\documentclass{ieeeaccess}

\usepackage[numbers]{natbib}
\usepackage{amsmath,amssymb,amsfonts}
\usepackage{algorithmic}
\usepackage{graphicx}
\usepackage{textcomp}
\usepackage{fontawesome5}
\usepackage{longtable}
\usepackage{changepage}
\usepackage[hidelinks]{hyperref}
\usepackage{tabularray}
\usepackage{bm}
\usepackage{array}
\usepackage{tcolorbox}
\usepackage[justification=centering]{caption}

\def\BibTeX{{\rm B\kern-.05em{\sc i\kern-.025em b}\kern-.08em
    T\kern-.1667em\lower.7ex\hbox{E}\kern-.125emX}}

\begin{document}

\title{Ethical software requirements from user reviews: A systematic literature review}
\author{\uppercase{Aakash Sorathiya}\authorrefmark{1}, and \uppercase{Gouri Ginde}\authorrefmark{2}}

\address[1]{Department of Electrical and Software Engineering, University of Calgary, Calgary, Canada (e-mail: aakash.sorathiya@ucalgary.ca)}
\address[2]{Department of Electrical and Software Engineering, University of Calgary, Calgary, Canada (e-mail: gouri.ginde@ucalgary.ca)}

\begin{abstract}
\\
    \textbf{Context}: Growing focus on ethics within software engineering (SE), primarily due to the significant reliance of individuals' lives on software and the consequential social and ethical considerations that impact both people and society has brought focus on ethical software requirements identification and elicitation. User safety, privacy, and security concerns are of prime importance while developing software due to the widespread use of software across healthcare, education, and business domains. Thus, identifying and elicitating ethical software requirements from app user reviews, focusing on various aspects such as privacy, security, accountability, accessibility, transparency, fairness, safety, and social solidarity, are essential for developing trustworthy software solutions.\\
    \textbf{Objective}: This systematic literature review (SLR) aims to identify and analyze existing ethical requirements identification and elicitation techniques in the Context of the formulated research questions. \\
    \textbf{Method}:  We conducted an SLR based on Kitchenham et al's methodology. We identified and selected 47 primary articles for this study based on a predefined search protocol. Of these, 28 were conference articles, 12 were journal articles, 1 were workshop articles, and 6 were preprint (Arxiv) articles.\\
    \textbf{Result}: Ethical requirements gathering has recently driven drastic interest in the research community due to the rise of Machine Learning (ML) and Artificial Intelligence(AI) based approaches utilized in decision-making (recommendation, classification) within software applications. This SLR provides an overview of ethical requirements identification techniques and the implications of extracting and addressing them. This study also reports the data sources used for analyzing user reviews.\\
    \textbf{Conclusion}: This SLR provides an understanding of the ethical software requirements and underscores the importance of user reviews in developing trustworthy software. The findings can also help inform future research and guide software engineers or researchers in addressing software ethical requirements.
\end{abstract}

\begin{keywords}
Ethics; Software; Software requirements; Requirements engineering; User reviews; SLR;
\end{keywords}

\maketitle

\section{Introduction} \label{intro}
Ethics, a subdivision of philosophy, addresses questions concerning human values and ethical principles through the clarification of notions such as right/good and wrong/bad \cite{capurro1988information}, \cite{yucel2009globalization}, \cite{paul2013thinker}, \cite{veatch1977case}. Today, the study of ethics is prevalent across a diverse range of disciplines such as healthcare and business \cite{chung2008classification} due to its widespread implications on the lives of the masses. Ethics, in this context, refers to the moral considerations and responsibilities associated with creating software that affects individuals and society. Ethical principles aid software engineers in upholding the welfare, justice, and safety of users and society, guiding their activities and decisions throughout the software design and development process \cite{Gotterbarn2002SoftwareEE}.

Software can either uphold, support, weaken, or compromise the fundamental values shared by individuals and society. Ethical shortcomings in such software may result in significant repercussions, such as privacy violations, discriminatory practices, dissemination of false information, and adverse effects on individuals or the community as a whole \cite{wong2010recent}. An exemplary instance of this can be identified in the Volkswagen Turbo Diesel Injection (TDI) models produced between 2009 and 2015, during which Volkswagen manipulated the emissions testing system by engineering their diesel vehicles to emit lower levels of pollutants than they actually did \cite{mansouri2016case}. As a further illustration, consider robotics and autonomous software. These technologies have the potential to give rise to ethical concerns through the displacement of particular workers, resulting in adverse impacts on individuals' employment. Additionally, they may inadvertently introduce bias in decision-making processes, such as those stemming from ethnically skewed training data, posing risks to specific population segments \cite{winfield2019ethical}. Therefore, it is important to address ethical issues conscientiously to guarantee that the software produced not only fulfills functional requirements but also complies with moral and societal norms that protect the well-being of the wider community.

When addressing ethical considerations, there are issues that one needs to take into account: (i) ethics is context-dependent, (ii) ethics is not a clear matter of right or wrong, good or bad, or white or black, and (iii) ethical issues are not easy to recognize and deal with \cite{alidoosti2022ethics}. Ethical principles can guide SE, but simply following a set of rules and standards (e.g., the ACM/IEEE Software Engineering Code of Ethics \cite{gotterbarn1997software}) is insufficient because ethical aspects are not easily tested like functional requirements. Therefore, software engineers need explicit and systematic approaches that consider human and non-technical aspects from the early phases of software creation, such as software design \cite{thomson2001ethics} \cite{turilli2008ethics} \cite{berenbach2009professional} or requirements engineering (RE) \cite{biable2023proposed} \cite{biable2022ethical} \cite{hosseini2016foundations}.

RE is a critical stage in the software development life cycle (SDLC), as its success directly impacts the success of software projects \cite{shah2014review} \cite{mudduluru2016value}. Even if a software system is implemented correctly, it will fail if it does not meet the necessary requirements \cite{mudduluru2016value}. RE is a crucial yet challenging phase of SDLC \cite{shah2014review} \cite{mudduluru2016value} as it determines exactly what needs to be developed and ensures the specification and validation of detailed technical and non-technical requirements \cite{mudduluru2016value}. Similarly, this stage is particularly challenging when it comes to addressing ethical issues. In reality, ethical considerations are often overlooked during RE \cite{shah2014review}. This is because software engineers usually act in isolation from the software users \cite{begier2010users} and they are generally culturally unfamiliar with some or most of their users \cite{Vallor2013AnIT}. Thus, software engineers should take into consideration their users' feedback and assess what is at stack for them during the RE phase. User feedback, in the form of reviews, contains a wealth of information about user experience and expectations \cite{genc2017systematic}. User reviews provide insights into real-life experiences and perspectives, shedding light on ethical concerns that might not have been evident in the RE stage of the SDLC. They can uncover algorithmic biases, privacy issues, or accessibility concerns that were overlooked at the beginning. However, it is difficult for an individual to read all the reviews and reach an informed decision due to the ever-growing amount of textual review data \cite{genc2017systematic}. Hence, over the last several years, various techniques and automated systems have been proposed to mine, analyze, and extract users' ethical concerns and sentiments from online reviews.

Despite research in the area of ethics and technology, there is insufficient knowledge on how to extract ethical concerns from user reviews and how to translate them into software requirements.\textbf{ This SLR is focused on determining} the research conducted so far regarding the utilization of user reviews to address ethical concerns. \textbf{The goal of this study }is to identify existing gaps in current research and to reveal potential future research directions that can enhance ethical considerations in the RE process.

The rest of this paper is organized as follows. Background information is covered in Section \ref{back}, discussing key concepts and the need for conducting this SLR. Section \ref{method} outlines the research methodology, detailing research questions, search strategy, research process, selection criteria, data extraction strategy, and data synthesis. Section \ref{result} presents the findings of the SLR, while Section \ref{discuss} delves into a discussion of the SLR. Section \ref{limit} describes some limitations and Section \ref{conclude} concludes the paper.

\section{Background} \label{back}
This section presents essential background and foundational concepts. Section \ref{cad} offers essential definitions related to SE and ethics, Section \ref{es} describes problems with traditional software and the need for ethical software, Section \ref{re} presents requirements of ethical software, and Section \ref{slr} summarizes the need for this SLR.

\subsection{\textbf{Concepts and definitions}} \label{cad}

\underline{\textbf{Software Engineering}}: Currently, the field of SE has become one of the top professions globally\cite{lee2014conceptual}. The significance of value-driven software has grown substantially due to the advancements in SE. This shift is crucial to meet the increasing demands of businesses, stakeholders, market conditions, competitiveness, and the profitability of organizations \cite{zakaria2015examining}. Whittle et al. \cite{whittle2019case} emphasize the importance of designing software with a focus on value by examining real-life projects. They argue that value-driven software development is key in incorporating human values into software products. While values and ethics may differ \cite{whittle2019case}, there is a strong connection between human values and ethical values \cite{evans2019ethics}. Therefore, software engineers must consider ethical values in their work to improve the quality of the products they create \cite{lee2021there}.

\underline{\textbf{Requirements Engineering}}: RE is a crucial aspect of engineering that concentrates on the specific objectives, functions, and limitations of a system \cite{laplante2022requirements}. It involves identifying, examining, defining, confirming, and upkeeping the system's requirements \cite{shah2014review} \cite{laplante2022requirements}. This includes understanding stakeholders' needs, grasping the requirements' context, representing them clearly, and handling all aspects of the requirements. Neglecting to identify the correct needs can result in flaws in specifications and other system components \cite{darwish2016requirements}. Poorly defined requirements can lead to various challenging issues that affect both direct and indirect stakeholders of software. RE provides appropriate tools for recognizing and evaluating users' genuine needs, assessing feasibility, clearly specifying solutions, validating requirements, managing them effectively, and selecting a practical solution for implementation \cite{roger2015software}. It encompasses several sub-processes, with the main five being requirements elicitation, requirements analysis and negotiation, requirements specification, requirements validation, and requirements management \cite{laplante2022requirements} \cite{darwish2016requirements} \cite{roger2015software} \cite{sommerville2011software} \cite{babar2015stakemeter}.

\underline{\textbf{Ethics}}: The term ``ethics" is derived from the Greek term ``ethos", which means character \cite{thiroux2016ethics}. In the realm of academia, ethics is primarily associated with philosophy. It involves examining various concepts like good, bad, right, and wrong as they relate to human behavior from a moral perspective \cite{thomson2001ethics}. Ethics deals with the guidelines for and exploration of moral principles, as well as the comprehension and acceptance of moral values within a specific setting or environment that needs to be clearly defined \cite{field2020moral}. In engineering, ethics are defined as “the study of moral issues and decisions confronting individuals and organizations involved in engineering” \cite{lynch2017engineering}. Engineering ethics are mainly inclined to the professional ethics of software engineers and seek to identify and question engineering standards and the ways that those standards are performed in specific circumstances \cite{biable2022ethical}.

\underline{\textbf{Ethical Values}}: Values have long been a fundamental concept in the social sciences, playing a crucial role in various scientific fields \cite{alidoosti2022ethics}. Different definitions exist for values, and some of the commonly used ones are outlined here. Kluckhohn et al. \cite{kluckhohn1951values} define value as “a conception, explicit or implicit, distinctive to an individual, or characteristic of a group, of the desirable which influences the selection from available modes, means, and ends of action”. Braithwaite et al. \cite{braithwaite1998consensus} view value as “principles for action covering abstract goals in life and modes of conduct that an individual or a collective deems preferable across various contexts and situations”. Schwartz et al. \cite{schwartz1994there} also see value as “a belief about desirable end states or modes of conduct that go beyond specific situations; it guides the selection or evaluation of behavior, people, and events; and is ranked in importance relative to other values to create a hierarchy of value priorities”. Additionally, Hutcheon et al. \cite{hutcheon1972value} argue that “values are not synonymous with ideals, norms, desired objects, or professed beliefs about the ‘good’, but rather are operational criteria for action”. Some examples of values include human welfare, privacy, security, trust, accessibility, fairness, and autonomy.

\underline{\textbf{User Reviews}}: User reviews are in-depth comments given by individuals regarding their interactions with a particular software. These reviews can be found on platforms like app stores, social media, and specialized review websites, offering important perspectives on the software's performance, ease of use, and possible ethical considerations \cite{genc2017systematic}. They not only assess user satisfaction but also bring attention to the wider social and ethical implications of the software.

\subsection{\textbf{Need for ethical software}} \label{es}
Software is rapidly changing every aspect of life, from online banking to healthcare management, education, and more \cite{diamandis2020future}. This widespread use of software brings many advantages but also raises ethical concerns. If software is not created and managed according to strict ethical standards, it can become risky and harmful. One such example is the Volkswagen emissions crisis \cite{mansouri2016case}. Similar software errors led to patients receiving lethal radiation dosages in the well-known instance of Therac-25, a radiation therapy equipment \cite{leveson1993investigation}. These incidents highlight how vitally important ethical considerations are while developing software. The significance of ethical software is further emphasized by issues related to privacy and data security. The Facebook-Cambridge Analytica data scandal, for example, demonstrated how user data can be abused without authorization, resulting in serious privacy and trust violations \cite{ur2019facebook}. Additionally, it has been shown that well-known health applications share user data with third parties without explicit consent, which raises serious privacy concerns \cite{grundy2019data}. Biased software algorithms can perpetuate and worsen social inequalities. In a notable case, a recruitment algorithm used by a major tech company was found to be biased against female candidates, reinforcing gender discrimination in hiring \cite{dastin2022amazon}. Furthermore, algorithmic wage discrimination has been identified in gig economy platforms, where software can lead to unfair pay gaps based on the demographics of the user \cite{dubal2023algorithmic}.

This evidence shows that not taking ethical considerations into account, particularly in the initial stages of developing software, i.e., RE, can result in creating software products that might not meet societal needs and could harm the general public. To mitigate this, ethical concerns must be considered in the RE process.

\subsection{\textbf{Requirements for ethical software}} \label{re}
RE consists of two types of requirements: functional requirements (FRs) and non-functional requirements (NFRs). FRs are centered on the key features that the system must provide as requested by the end-user, while NFRs focus on the system's performance and its adherence to specific standards or ethical values \cite{sajid2024addressing}. These NFRs are fundamental elements that contribute to the success of a system \cite{glinz2007non} and they account for ethical values also \cite{barn2016you}. The concept of Value Sensitive Design (VSD) \cite{friedman1996value} offers a robust framework for incorporating ethical values systematically. VSD is considered the most comprehensive method to account for human values throughout the design process \cite{winkler2021twenty} \cite{manders2011values} \cite{davis2015value} \cite{friedman2013value}. Another value structure, provided by Schwartz, recognizes 59 human values in 10 value categories in which each value category has different motivational goals \cite{schwartz1992universals} \cite{schwartz2012overview}. Moreover, Wright's proposed Ethical Impact Assessment (EIA) framework presents a structured approach to evaluating the ethical consequences of technology, guaranteeing that both potential risks and advantages are thoughtfully examined \cite{wright2011framework}. All these studies offer a sheer volume of ethical values and they are found to be converging around these eight main ethical requirements: \textit{privacy}, \textit{security}, \textit{accessibility}, \textit{accountability}, \textit{transparency}, \textit{fairness}, \textit{safety}, and \textit{social solidarity}. These requirements are explained below:

\textit{\textbf{Privacy}}: Privacy is the practice of protecting data from unauthorized use, access, and disclosure \cite{wright2011framework} \cite{friedman2013value} \cite{european-commission-2019}.

\textit{\textbf{Security}}: Security is the practice of safeguarding information and resources from unauthorized access, disclosure, destruction, or modification \cite{wright2011framework} \cite{schwartz1992universals}.

\textit{\textbf{Accessibility}}: Accessibility is the practice of designing software in a way that makes it accessible and easy to use for more people e.g. senior citizens or citizens with disabilities \cite{wright2011framework} \cite{friedman2013value}.

\textit{\textbf{Accountability}}: Accountability is an ability to explain the decisions and answers given by the software, which increases transparency and trust in the software \cite{european-commission-2019}. It also involves identifying and addressing the negative and positive consequences of the software and taking responsibility for it \cite{wright2011framework} \cite{friedman2013value}.

\textit{\textbf{Transparency}}: Transparency is the ability of software to provide an understanding of how it works and explain the decisions it makes \cite{wright2011framework} \cite{european-commission-2019}.

\textit{\textbf{Fairness}}: Fairness is the practice of ensuring that software should be able to make decisions without bias or prejudice towards any person or group \cite{wright2011framework} \cite{european-commission-2019} \cite{friedman2013value}.

\textit{\textbf{Safety}}: Safety is the practice of ensuring that software does not cause any physical or psychological harm to people \cite{schwartz1992universals} \cite{wright2011framework}.

\textit{\textbf{Social Solidarity}}: Social solidarity is the practice of ensuring that software does not have any negative impact on society like social isolation of any individual or substituting human contact \cite{wright2011framework} \cite{schwartz1992universals} \cite{friedman2013value}.

\textbf{This list provides essential requirements for designing ethical and trustworthy software and it also aligns with the guidelines provided by the EU \cite{european-commission-2019} for designing and developing trustworthy AI software.}

\subsection{\textbf{Need for SLR}} \label{slr}
Over the past decade, the SE field has seen a rise in attention towards ethics, leading to a focus on user reviews to understand ethical concerns. Despite an extensive review of relevant literature, no review studies were found in this particular area. Therefore, this work involves conducting an SLR to systematically examine and collect research findings to comprehend users' ethical concerns related to software. The objective of this review is to assist software engineers and researchers in gaining insight into the latest developments in understanding users' ethical concerns and the implications/guidelines to address them. This review also identifies the primary issues and challenges in the current literature and suggests directions for future research.

\begin{figure*}[t]
    \centering
    \includegraphics[width=1\linewidth]{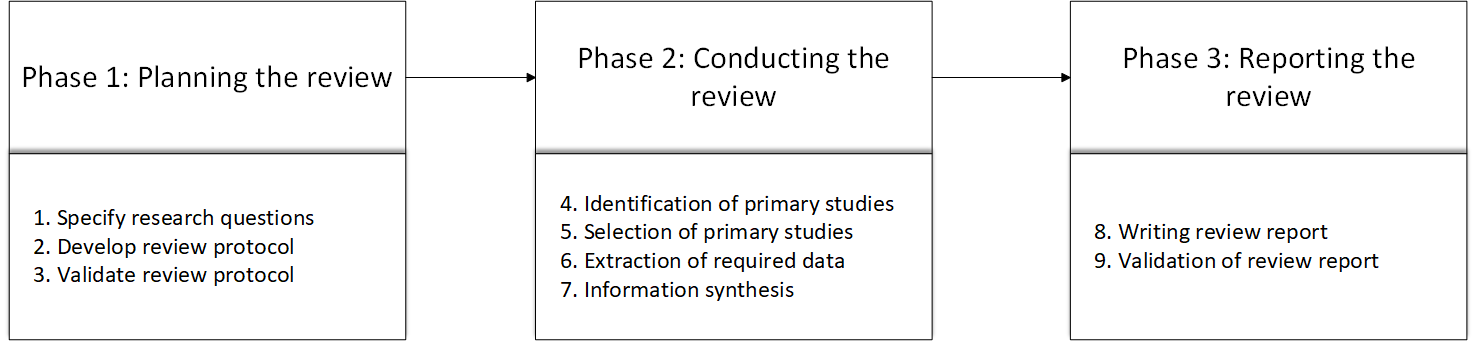}
    \caption{We follow the SLR process outlined by Kitchenham et al. \cite{keele2007guidelines} which has three phases to it.}
    \label{fig:slr_process}
\end{figure*}

\section{Research methodology} \label{method}
The SLR was carried out following the guidelines of Kitchenham et al. \cite{keele2007guidelines}. The SLR process involved three main phases: planning, conducting, and reporting the review as shown in Figure~\ref{fig:slr_process}. Each phase included specific tasks outlined in the below sections.

\subsection{\textbf{Planning}}
The planning stage made clear the precise goals of the SLR, which were to identify user review studies that concentrate on analyzing user ethical concerns, the methods and implications to assist software engineers and researchers, and any unresolved challenges. Furthermore, we outlined four research questions and the rationale behind them.

\subsubsection{\textbf{Research questions}}
\begin{itemize}
    \item \textbf{RQ1}: What software categories (gaming, health, education, etc.) are commonly examined for understanding users' ethical concerns?\\
    \textbf{Rationale}: Identifying software categories could guide future research efforts toward underexplored areas and prompt the development of new policies or guidelines to ensure that ethical standards are followed across all software categories.
    \item \textbf{RQ2}: What primary data sources have been commonly used in prior research to examine user reviews?\\ 
    \textbf{Rationale}: User reviews are a large corpus of unstructured data \cite{li2020user}, so it becomes very important to identify the relevant data source that is robust and suitable to examine ethical concerns. It also enables the aggregation of various data sources for more comprehensive and generalizable findings.
    \item \textbf{RQ3}: What specific techniques have been used to extract and analyze user reviews related to ethical concerns?\\ 
    \textbf{Rationale}: Over time, a variety of techniques and automated systems have been created to extract and analyze user opinions and sentiments from the large amount of unstructured data found in user reviews. This question seeks to investigate the specific methods and strategies that have been suggested and used to efficiently analyze user reviews related to ethical concerns.
    \item \textbf{RQ4}: What are the implications from the previous studies that can help software engineers address ethical concerns in the RE phase of the SDLC?\\ 
    \textbf{Rationale}: To address the ethical concerns of users they need to be translated into requirements that can be taken into consideration during the RE phase. By answering this question we aim to highlight the approaches and guidelines that can help software engineers to address users' ethical concerns and develop ethical and trustworthy software.
\end{itemize}

\subsubsection{\textbf{Development and validation of the review protocol}}
The review protocol outlines the necessary steps for conducting the literature review. By establishing a review protocol, potential biases among researchers are minimized, and parameters for source selection, search methods, quality standards, and synthesis techniques are set. This subsection elaborates on the specifics of the review protocol.

The following six digital libraries were searched for primary studies:
\begin{itemize}
    \item Scopus
    \item Web of Science (WoS)
    \item ScienceDirect
    \item ACM Digital Library (DL)
    \item IEEE Xplore
    \item ArXiv (non-peer-reviewed/work under progress/preprint literature)
\end{itemize}

The years covered by the search were between 2015-August 2024. 

Then, we defined the search query using the PICO method, as suggested by Kitchenham et al. \cite{keele2007guidelines}. The PICO method describes the population, intervention, comparison, and outcomes of the research questions to define a search query.

- Population: refers to the group of individuals, objects, or an application area being studied. In this case, the population is user reviews.

- Interventions: refers to the methodology, technology, or procedure being investigated. In this case, the intervention is the analysis of user reviews, specifically to identify and address ethical concerns.

- Comparison: refers to the methodology, technology, or procedure with which the previous intervention is compared. In this case, there is no comparison as only ethical concerns are analyzed from user reviews.

- Outcomes: refers to the overall outcome of the intervention. As we study how user reviews are analyzed to address ethical concerns, the findings identified are used as the outcome.

Using the above definitions, a search string was developed combining the sections with ``AND" statements as well as ``OR" statements between each word within the section. Using the keywords with possible synonyms the final query created was as follows: \textbf{(``user reviews" OR ``user feedback" OR ``user comments" OR ``user perspectives") AND (ethics OR ethical OR concerns)}

\textbf{Study selection procedure}: We systematically selected the primary studies by applying the following five steps:
\begin{enumerate}
    \item We examined the study titles to eliminate studies unrelated to the research focus.
    \item We evaluated the abstracts and keywords present in the remaining research investigations. In cases where the information in the abstracts or keywords was insufficient, an examination of the results and conclusion sections was conducted to ascertain the relevance of the study.
    \item The remaining studies were subjected to filtration based on the inclusion and exclusion criteria outlined in Table~\ref{tab:icec}. After this, we removed duplicated studies.
    \item We then applied single-step backward and forward snowball sampling from the references and citations of the filtered studies for any additional relevant studies.
    \item At the end, we assessed the quality of each study using a set of questions, shown in Table~\ref{tab:qe}. We evaluated all the studies and scored them against each question with 1 (``yes") or 0 (``no"). Then, a total score was assigned to each study by summing up the scores for all the questions; therefore each study could obtain a score ranging from 0 (very weak) to 5 (very strong). The studies with a total score of more than 2 were selected as primary studies. Note that QE5 was considered the minimum quality threshold; if this question received a negative response the study was discarded due to a lack of quality.
\end{enumerate}

\begin{table}[]
    \centering
    \caption{Inclusion and exclusion criteria}
    \label{tab:icec}
    \begin{tabular}{p{8cm}}
        \textbf{Inclusion criteria (IC)} \\
        \hline
        \\
        \textbf{IC1}. Articles published in a peer-reviewed outlet contained in ACM Digital Library, IEEE Xplore, Science Direct, Scopus, the Web of Science, or for grey (non-peer-reviewed) literature, ArXiv. \\
        \textbf{IC2}. Articles needed to be in English. \\
        \textbf{IC3}. Articles included the relevant search terms as previously defined  \\
        \textbf{IC4}. Articles that were published after 2014 (2015 or later)  \\
        \textbf{IC5}. Articles that describe users' concerns through the analysis of their reviews \\
        \hline
        \\
        \textbf{Exclusion criteria (EC)} \\
        \hline
        \\
        \textbf{EC1}. Articles that did not meet inclusion criteria. \\
        \textbf{EC2}. Articles that are not full-text available. \\
        \textbf{EC3}. Articles that focus on user reviews but only casually mention ethics. \\
        \textbf{EC4}. Articles that focus on ethical considerations of software but do not consider user reviews. \\
        \textbf{EC5}. Articles that are survey studies, but the questionnaire is not related to ethical concerns. \\
        \textbf{EC6}. Articles that discuss hardware/product/device issues and not software. \\
        \textbf{EC7}. Articles that do not analyze English language user reviews. \\
        \hline
        \\
    \end{tabular}
\end{table}

\begin{table}[]
    \centering
    \caption{Quality evaluation checklist}
    \label{tab:qe}
    \begin{tabular}{p{8cm}}
        \textbf{Quality evaluation checklist (QE)} \\
        \hline
        \\
        \textbf{QE1}. Does the study's aim focus on uncovering ethical concerns from user reviews? \\
        \textbf{QE2}. Does the study explain how and from where the data was collected? \\
        \textbf{QE3}. Is the methodology explained appropriately? Does it show the approach to identifying ethical concerns from user reviews? \\
        \textbf{QE4}. Does the study discuss in detail the ethical concerns of users? Does it provide any implications for software engineers? \\
        \textbf{QE5}. If the study includes other data like software policies in analyzing software’s ethical landscape, does the user review data play a significant role in it? \\
        \\
    \end{tabular}
\end{table}

\begin{figure*}[h!]
    \centering
    \includegraphics[width=1\linewidth]{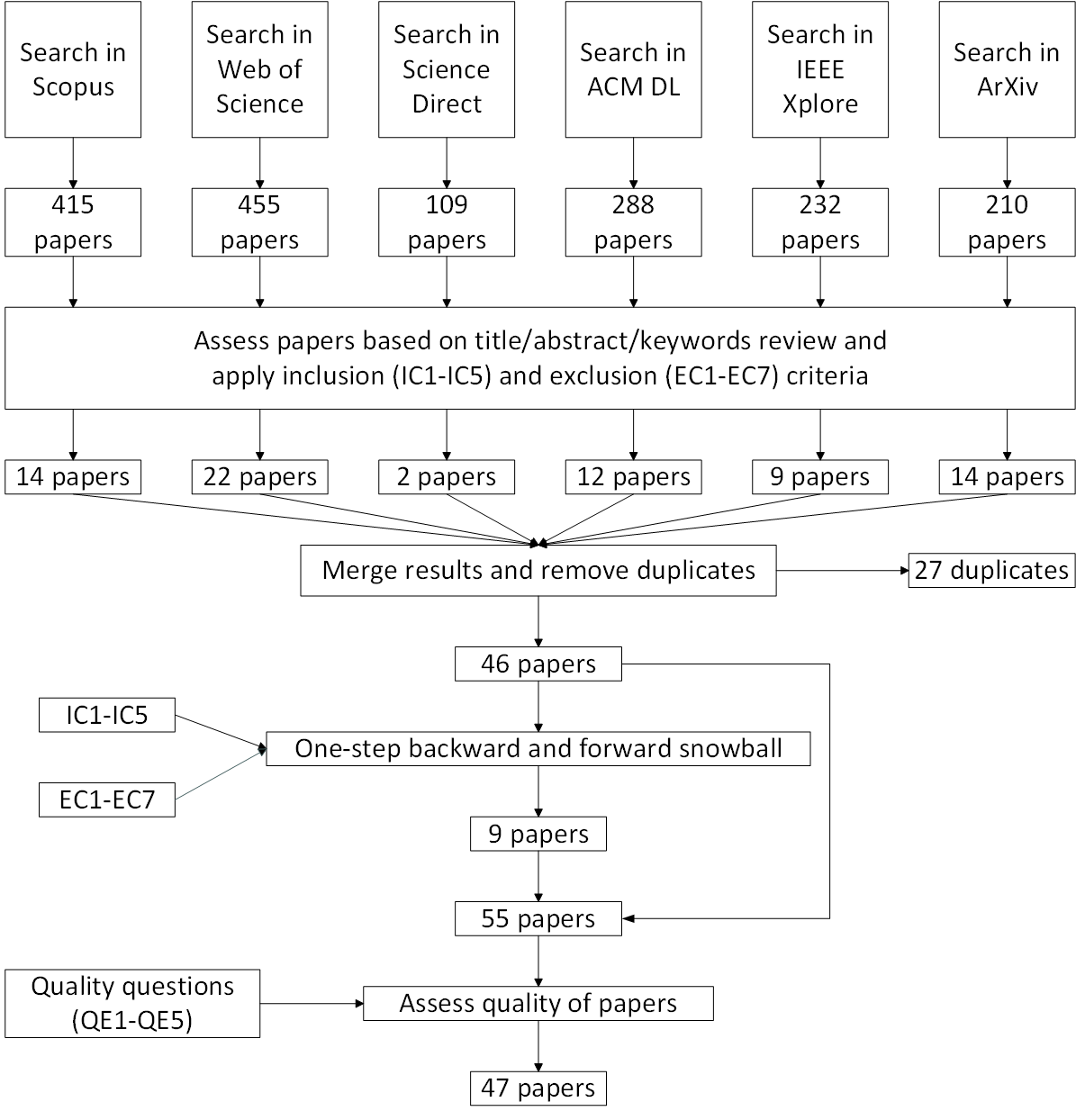}
    \caption{Details of study selection process}
    \label{fig:search}
\end{figure*}

\subsection{\textbf{Conducting the review}}

\subsubsection{\textbf{Identification and selection of the primary studies}}
Figure \ref{fig:search} shows the details of the study selection procedure. By adhering to the search strategy detailed in the previous section, the digital libraries were explored and the articles were retrieved. Initially, 1709 articles were found, with some being duplicates due to overlapping sources in the repositories. These articles underwent a rigorous screening process following steps 1, 2, and 3 of the study selection procedure. This resulted in 74 articles, out of which 27 duplicate articles were identified and removed, leaving 47 unique articles. Then step 4 was carried out which yielded an additional 8 articles. In total, 55 articles were then assessed for quality as mentioned in step 5. After this final evaluation, 47 articles were ultimately selected as primary studies for this SLR which are listed in Appendix \ref{appendixA}.

\subsubsection{\textbf{Data extraction}}
We used the data extraction form in Table~\ref{tab:extraction} to extract data from the 47 primary studies. Refer Appendix \ref{appendixB}.

\begin{table}[]
    \caption{Data extraction form}
    \label{tab:extraction}
    \begin{tabular}{ll}
         \textbf{Data item} & \textbf{Description} \\
         \hline
         \\
         Identifier & Reference number given to the study \\
         Bibliography & Authors, year of publication, title, source \\
         Type of study & Journal/conference/technical report/etc. \\
         Author affiliation & Academic/Industrial/Collaboration \\
         Study aims & Aims or goals of the study \\
         Software dataset & Categories and number of software \\
         User reviews dataset & Source and size of user reviews data \\
         Methods and techniques & Algorithms, models and measures \\
         Implications/Guidelines & Suggestion or tools to address user concerns \\
         Ethical concerns & Concerns related to ethical requirements \\
         \\
    \end{tabular}
\end{table}

\subsubsection{\textbf{Information synthesis}}
We reviewed 47 selected primary studies, highlighting the techniques and outcomes that were replicated. Knowledge gaps and open challenges were also documented and are presented in the discussion section.

\subsection{\textbf{Reporting the review}}
The data obtained from the primary studies were employed to address the four research questions. We adhered closely to Kitchenham's \cite{keele2007guidelines} guidelines when presenting the results.

\section{Results} \label{result}
In this section, we provide a broad summary of the trends in publications and the outcomes related to the research questions.

\subsection{\textbf{Publication Trends}}
In this section, we aim to discover the trends in the primary studies based on publication year, publication venue, venue type, authors' collaboration, and citations.

\begin{figure*}[h!]
    \centering
    \includegraphics[width=1\linewidth]{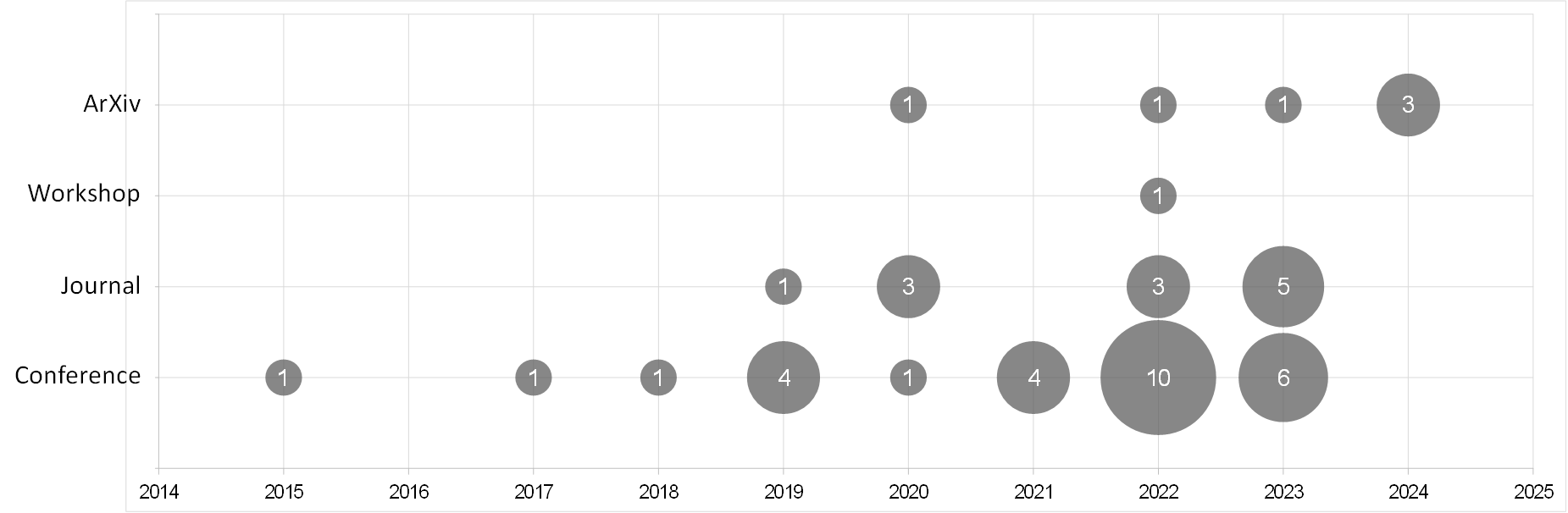}
    \caption{Number of primary studies based on publication year and venue type}
    \label{fig:trends}
\end{figure*}

In recent decades, there has been a noticeable but modest increase in the number of publications focusing on ethical considerations from user reviews. The relationship between the publication year and venue type of these studies is depicted in Figure~\ref{fig:trends}, with the size of bubbles indicating the corresponding number of studies. The primary studies included in this analysis range from the years 2015 to 2024, revealing a gradual rise in publications in more recent years. 

The predominant venue types targeted by these studies are conferences and journals. Notably, conferences and journals account for the majority of the selected studies, representing 57\% (28 out of 47 studies) and 28\% (12 out of 47 studies) respectively. Certain publication venues are particularly prolific in publishing articles related to this area. An overview of the primary studies, including publication venues and venue types, can be found in Table~\ref{tab:venue} of Appendix \ref{appendixC}. This table illustrates that the 47 primary studies are distributed across 29 different publication venues including the pre-print publication venue ArXiv. The majority of these venues have published one to two articles on the subject. Among the preferred venues is the ``CHI Conference on Human Factors in Computing Systems" which has published five articles. Additionally, six articles are pre-printed and published on ArXiv. 

Table~\ref{tab:aff} shows authors' affiliation with their respective studies, where thirty-nine studies are academic, six studies are collaboration, and just two studies are industry-affiliated.

\begin{figure*}[h!]
    \centering
    \includegraphics[width=1\linewidth]{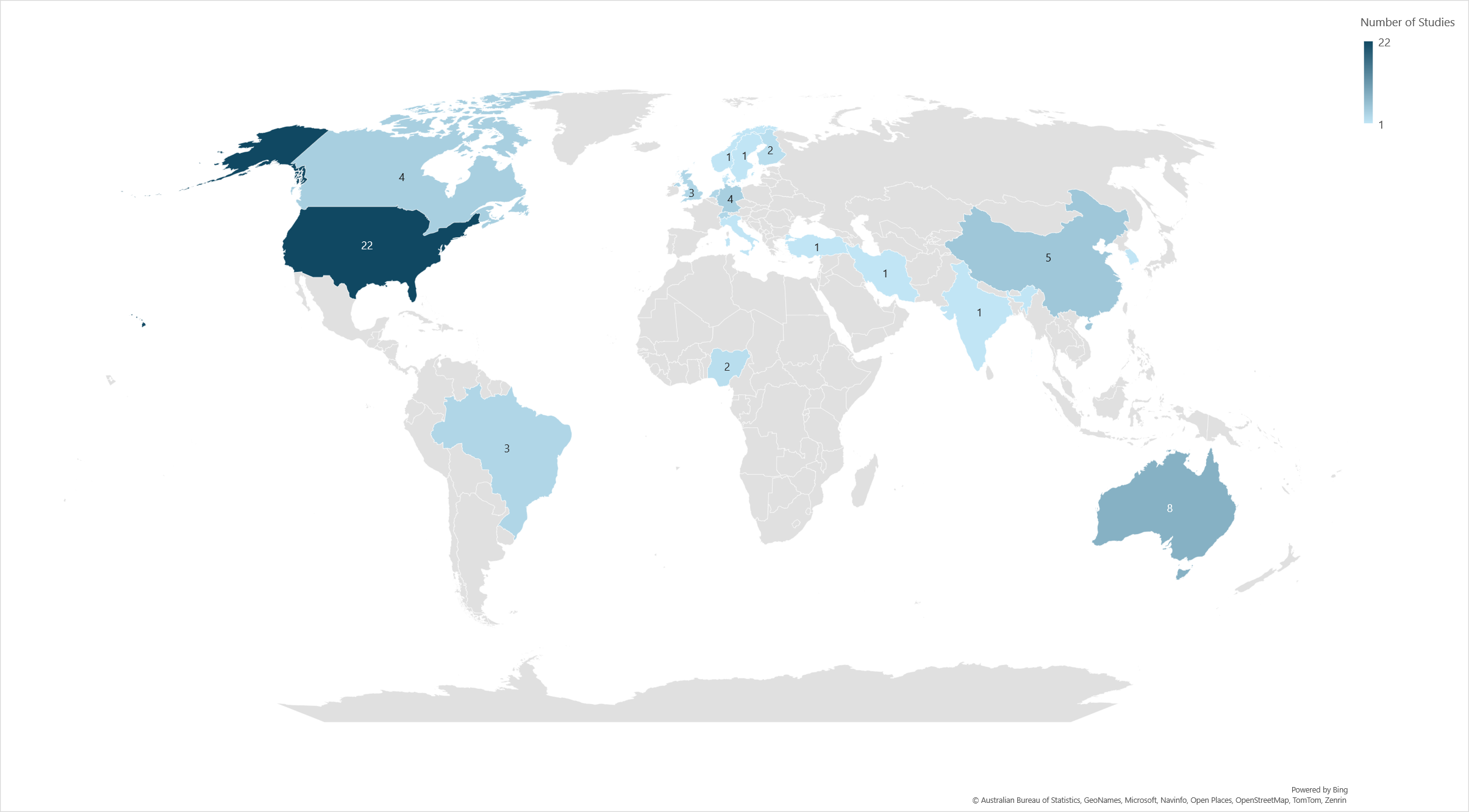}
    \caption{The geographical distribution of the papers published by teams on the subject.}
    \label{fig:map}
\end{figure*}

Map in Figure \ref{fig:map} depicts the graphical distribution of teams, with the highest number of studies published by the authors from the United States team (22 studies). Note that we included the nationality of all authors in the analysis. Among all studies, 16 studies [S2, S17, S18, S26, S27, S31, S33, S35, S36, S37, S38, S41, S42, S43, S44, S47] had authors from more than one country.

We also analyzed the impact of studies in terms of their citation count as shown in Table \ref{tab:cites}. The average number of citations is 13.51 with studies S3 and S30 receiving the highest number of citations with a count of 65 and 82, respectively. Note that the citation count is fetched using SemanticScholar API and it may change over time.

\begin{table}
    \centering
    \caption{Primary studies citation ranges.}
    \label{tab:cites}
    \begin{tabular}{lp{1.7cm}p{3.8cm}}
        \textbf{Citation range} & \textbf{Number of primary studies} & \textbf{Primary studies} \\
        \hline
        0 & 4 & S8, S35, S34, S36 \\
        1-5 & 18 & S2, S7, S12, S16, S44, S32, S33, S41, S40, S46, S18, S11, S29, S10, S20, S22, S43, S37 \\
        6-10 & 7 & S9, S28, S4, S5, S26, S45, S38 \\
        11-20 & 6 & S1, S6, S13, S47, S15, S17 \\
        21-30 & 4 & S31, S23, S25, S24 \\
        31-40 & 2 & S19, S39, S27 \\
        41-50 & 2 & S21, S14 \\
        65-85 & 2 & S3, S30 \\
    \end{tabular}
\end{table}

\subsection{\textbf{RQ1: Software categories}}
Table~\ref{tab:categories} outlines the software categories explored with their respective references. The software categories mentioned in the primary studies were grouped based on the guidelines provided by the Google Play Console\footnote{\url{https://support.google.com/googleplay/android-developer/answer/9859673?hl=en\#zippy=}}. Most of the studies (22 out of 47) didn't focus on a specific software category but instead carried out generic analyses. Among others, the software categories explored range from health-related software to graphic designing tools. The health category, which includes fitness, mental health, depression, teletherapy, and COVID-19 related software, appears to be the most studied area with 9 studies. The second most prominent category is the sharing economy, covering food delivery, ride-sharing, lodging, and freelancing platforms. This category is explored in 3 studies. Both finance (including investing and mobile payment) and social (including social VR and social networks) categories have been explored in 3 and 4 studies respectively. The gaming category, which includes metaverse software, is covered in 3 studies. Education and graphic designing appear to be less explored, with only one study each. Study S24 span across 3 different categories: health, finance, and sharing economy.

\begin{table}[h]
    \centering
    \caption{Table shows the software categories explored by the primary studies.}
    \label{tab:categories}
    \begin{tabular}{p{3.8cm}p{3.8cm}}
         \textbf{Software category} & \textbf{Studies}  \\
         \hline
         \hline
         Health (Fitness, Mental Health, Depression, Teletherapy, Covid-19) & S1 S4 S6 S10 S12 S13 S28 S30 S31 \\
         \hline
         Sharing Economy (Food Delivery, Ride-Sharing, Lodging, Freelancing) & S4 S9 S16 \\
         \hline
         Finance (Investing, Mobile Payment) & S4 S7 S22 \\
         \hline
         Gaming (Metaverse) & S8 S11 S15 \\
         \hline
         Social (Social VR, Social Network) & S2 S29 S35 S43 \\
         \hline
         Education & S23 \\
         \hline
         Graphic Designing & S34 \\
         \hline
         General (No specific category) & S3 S5 S14 S17 S18 S19 S20 S21 S24 S25 S26 S27 S32 S33 S36 S37 S38 S39 S40 S41 S42 S44 S45 S46 S47 \\
         \\
    \end{tabular}
\end{table}

\begin{tcolorbox}[arc=0mm,width=\columnwidth,
    top=1mm,left=1mm,  right=1mm, bottom=1mm,
    boxrule=1pt] 
Summary of RQ1: Most of the studies have a generic focus and among others health category is found to be the most explored one. Sharing economy, finance, social, and gaming categories also receive significant attention. While the research covers a broad spectrum of categories, some areas like education and graphic designing are less explored.
\end{tcolorbox}

\subsection{\textbf{RQ2: Data source for user reviews}}
Table~\ref{tab:datasource} shows the list of data sources used by the primary studies to analyze user reviews. The Google Play Store is by far the most commonly used data source, utilized in 37 studies. The second most popular data source is the Apple App Store, used in 10 studies. Reddit and Twitter are also used as data sources, with Reddit appearing in 4 studies and Twitter in 2. This indicates researchers are looking beyond app stores to gather user reviews. Some studies use more specialized platforms like Meta Quest for VR-related software, Steam for desktop gaming software, Chrome Web Store for browser extensions, and Amazon App Store as an alternative app store. One study explores an online forum (DPReview) showing the potential towards topic-specific communities. Few studies have also used multiple data sources. These results show the diversity in data sources considered by the primary studies.

\begin{table}[h]
    \centering
    \caption{Table shows the data sources used by primary studies for analyzing user reviews}
    \label{tab:datasource}
    \begin{tabular}{p{3.8cm}p{3.8cm}}
         \textbf{Data source} & \textbf{Studies} \\
         \hline
         \hline
         Google Play Store & S1 S3 S4 S6 S7 S8 S10 S11 S12 S13 S14 S15 S17 S18 S19 S20 S21 S22 S24 S25 S26 S27 S28 S30 S31 S32 S33 S36 S37 S38 S39 S40 S41 S42 S44 S46 S47 \\
         \hline
         Apple App Store & S1 S4 S6 S12 S13 S16 S28 S30 S35 S43 \\
         \hline
         Amazon App Store & S45 \\
         \hline
         Meta Quest & S2 \\
         \hline
         Reddit & S5 S29 S33 S34 \\
         \hline
         Twitter & S9 S33 \\
         \hline
         Steam & S15 \\
         \hline
         Chrome Web Store & S23 \\
         \hline
         Online Forum (DPReview) & S34 \\
         \hline
         GitHub & S42 \\
         \\
    \end{tabular}
\end{table}

\begin{tcolorbox}[arc=0mm,width=\columnwidth,
    top=1mm,left=1mm,  right=1mm, bottom=1mm,
    boxrule=1pt] 
Summary of RQ2: This analysis of data sources for user reviews reveals a strong preference for mobile app stores, with Google Play Store dominating (37 studies) followed by Apple App Store (10 studies). While mobile app stores are the primary focus, a few of the studies also utilize social media (Reddit, Twitter) and niche platforms (Meta Quest, Steam, Chrome Web Store) to gather diverse user reviews. Some studies combine multiple sources for a more comprehensive analysis.
\end{tcolorbox}

\subsection{\textbf{RQ3: Techniques to extract and analyze users' concerns}}
Here we discuss in detail the techniques used by primary studies to extract and analyze users' concerns related to ethical requirements which are presented in Section \ref{re}. Table \ref{tab:methods} of Appendix \ref{appendixB} summarizes the list of various techniques employed by primary studies and Table \ref{tab:concerns} shows the categorization of primary studies into the ethical requirements they address. Three main stages are observed in the whole process: identifying concern-related reviews, automated classification of reviews, and analysis of concern-related reviews.

Study S1 utilizes purposive sampling for the identification of relevant user reviews from app stores. Analysis of reviews is based on their helpfulness and critical perspective, as determined by filters within the app store platform. Exclusion criteria are applied to reviews lacking substantial content, such as those comprising solely of emojis or brief statements, to guarantee that the analysis encompasses detailed user experiences and ethical considerations. Thematic analysis \cite{braun2012thematic} using ATLAS.ti \cite{smit2002atlas} is employed to delve into user experiences and ethical concerns. This approach includes categorizing reviews as positive, negative, or ambivalent, and developing inductive codes to grasp the substance and context of user opinions. The final code inventory is organized into themes that mirror ethical principles and user experiences.

Study S2 analyses user reviews from social VR communities and reports concerns, like harassment and violent language, related to safety. It performs thematic analysis \cite{braun2012thematic} on reviews that were extracted from the Meta Quest platform. 

Studies S3, S20, S21, S24, and S40 suggest various techniques to automatically classify privacy and security-related user reviews. Study S3 presents AUTOREB, a system crafted to deduce privacy and security-related concerns from user reviews automatically. This system utilizes a feature augmentation strategy that delves into the concept of ``relevant feedback," which takes into account the associations and co-occurrences within various reviews. This technique proves especially beneficial for addressing the concise and often inadequately articulated nature of user reviews, aiding in capturing adequate semantics to comprehend the concerns described by users. AUTOREB employs a machine learning classifier, Sparse SVM \cite{cotter2013learning}, for the automated classification of user reviews into the four predefined categories of privacy and security behaviors. To enhance result reliability, the system aggregates these behaviors from the review level to the application level by leveraging crowdsourcing methods. It then ranks the app based on the number of behavior violations. 

Study S20 uses keyword search and manual analysis to prepare a training set for automation and then employs an artificial neural network (ANN) \cite{zou2009overview} to automatically classify user reviews with privacy and security concerns. Study S21 uses a keyword search with permission manifest files for all applications and creates a vector of privacy and security threats and then uses cosine similarity \cite{salton1988term} to identify and label reviews related to privacy and security.
It then explores Support Vector Machine (SVM) \cite{yue2003svm}, Logistic Regression (LR) \cite{lavalley2008logistic}, Decision Trees (DT) \cite{kingsford2008decision}, Random Forest (RF) \cite{rigatti2017random}, and K-Nearest Neighbours (KNN) \cite{guo2003knn} models for automated classification. 

Study S24 also uses keyword search to identify relevant reviews and then manually creates the Semantic Dependency Graph (SDG) \cite{de2006generating} to extract (misbehavior-aspect-opinion)  triples from the review sentences. It then employs the LR model \cite{lavalley2008logistic} with Bag-of-Words (BoW) \cite{zhang2010understanding} and TF-IDF \cite{yun2005improved} combination to automatically mine privacy and security-related reviews. Study S40 uses keyword search to identify relevant reviews and performs manual labeling to create training data for automated classification. It explores Naive Bayes (NB) \cite{rish2001empirical}, KNN \cite{guo2003knn}, Single-Layer averaged Perceptron (SLP) \cite{raudys1998evolution}, and SVM \cite{yue2003svm} models to automate the classification. Additionally, it analyzes the network traffic of apps to search for the presence of two types of Personal Information Identifiers (PII): data used to geolocate the user and persistent identifiers for tracking, to understand the security and privacy measures of an app.

Study S31 performs a multi-lateral analysis of Android apps to address privacy and security concerns. It uses the model from study S21 to automatically identify relevant user reviews and it also considers permission manifest files, privacy policies, and permission usage logs. It then manually identifies ten dangerous permissions from permission manifest files, checks if it is mentioned in the privacy policy, and if it is used by the app when idle. Based on this analysis it then ranks the app.

Studies S4, S35, and S43 propose unsupervised approaches for summarizing privacy and security-related concerns in user reviews of mobile apps. Study S4 uses keyword search with the initial seed of \{privacy, private, security\} terms and expands this set for domain-specific reviews by manual analysis. Then it uses the variant of the Hybrid TF-IDF \cite{yun2005improved} algorithm with GloVe \cite{pennington2014glove} embeddings of the identified keywords to summarize the reviews. This approach is shown to perform better than traditional Hybrid TF-IDF \cite{yun2005improved} and LDA \cite{jelodar2019latent}. Study S35 calculates the cosine similarity \cite{salton1988term} between privacy policies and user reviews to identify relevant reviews. To compute document embeddings for this task, it explores pre-trained sentence transformers\footnote{\url{https://www.sbert.net/docs/sentence_transformer/pretrained_models.html}}: all-mpnet, all-MiniLM, multi-qa-mpnet, and multi-qa-MiniLM. It then manually labels the reviews and explores LR \cite{lavalley2008logistic}, Linear SVM \cite{yue2003svm}, RF \cite{rigatti2017random}, Gradient Boosting DT (GBT) \cite{friedman2001greedy}, fastText \cite{joulin2016bag}, and BERT \cite{devlin2018bert} for automated classification. As a last step, it explores BERTopic \cite{grootendorst2022bertopic} with K-means algorithm \cite{ahmed2020k} to summarize the privacy and security-related reviews. 

Study S43 also uses privacy policies to identify relevant reviews but it specifically considers only collection, disclosure, retention, and usage-related privacy statements. It then employs SentenceTransformers \cite{reimers2019sentence} to encode the privacy statements and user reviews and calculates the cosine similarity \cite{salton1988term} to select the relevant reviews. As a next step, it uses LDA \cite{jelodar2019latent} to model identified reviews into the topics aligning with privacy statements and analyze the gaps between end users' perspectives and privacy policies.

Studies S5, S17, S27, S45, and S47 address privacy concerns and suggest multiple techniques for automated classification. Study S5 analyzes Reddit posts by using keyword search with a set of \{GDPR, CCPA, privacy, leak, consent, unlawful\} keywords to identify and manually label relevant reviews. It then uses the Universal Sentence Encoder (USE) \cite{cer2018universal} classifier with SMOTE \cite{chawla2002smote} to automatically classify concern-related reviews. As a next step, it summarizes concern-related reviews into nine main themes using the K-means algorithm \cite{ahmed2020k} with Distil-BERT \cite{sanh2019distilbert} for embedding the data and UMAP \cite{mcinnes2018umap} for dimension reduction. Study S17 investigates multiple techniques for addressing privacy-related concerns from user reviews. First, it utilizes a curated list of privacy-related n-grams transformed into regex to identify potential privacy reviews and create a training set for automated classification. It explores three models: USE \cite{cer2018universal}, BERT (Vanilla) \cite{devlin2018bert}, and BERT-SST (Sentiment-aware BERT) S17, for automated classification tasks. It then utilizes the ensemble model, which combines the results of the above three models. Additionally, it uses a K-means algorithm \cite{ahmed2020k} with silhouette score to summarize the underlying privacy themes within user reviews. 

Study S27 investigates Android run-time permissions through user reviews. It uses keyword search with the initial set of \{permission, privacy, consent\} keywords to identify relevant reviews and manually label the reviews into ten permission-related categories to create training data for automated classification. It then explores SVM \cite{yue2003svm}, NB \cite{rish2001empirical}, Classification and Regression Tree (CART) \cite{breiman2017classification}, and Maximum Entropy (ME) \cite{torgo2011data} models to automate the classification of privacy-related reviews. Study S45 uses random sampling, keyword sampling, corrective sampling, and manual analysis to identify and label the relevant reviews and afterward, it employs LR \cite{lavalley2008logistic} for automated classification. Additionally, it performs sentiment analysis and manually analyzes the most engaged reviews to compare privacy and security-related reviews with non-privacy and security-related reviews. Study S47 first extracts privacy concepts from privacy taxonomies, creates thirty-five hypotheses from it and then leverages natural language inference (NLI) \cite{maccartney2009natural} to identify relevant reviews. It uses the T5-11B \cite{raffel2020exploring} model for automated classification of reviews. Next, it manually annotates the reviews with a set of issues occurring in the review and trains the T5-11B model to summarize the privacy-related reviews into the type of issues. It then groups similar issues into one theme using the leader algorithm \cite{hartigan1975clustering} and again trains a T5-11B model to summarize reviews into high-level themes. Additionally, it extracts the emotions and quality of the reviews for understanding the informative reviews with their emotions present in various high-level themes.

Studies S6, S10, S12, and S30 analyze mobile mental health apps to identify and analyze user concerns. Study S12 analyzes only apps related to the teletherapy sub-category. All the studies perform the manual analysis to identify relevant reviews and employ thematic analysis \cite{braun2012thematic} to identify various aspects of apps from user reviews. Study S10 employs auditors to use apps for some time to validate the identified themes.

Studies S7 and S22 address user concerns by systematically analyzing reviews of mobile payment apps. Both studies use keyword search and sentiment analysis to identify relevant reviews and employ thematic analysis \cite{braun2012thematic} to analyze the reviews manually. Additionally, study S7 explores tools such as MobSF \cite{abraham2016mobile}, AndroBugs \cite{lin2015androbugs}, and RiskInDroid \cite{merlo2017riskindroid} to analyze APKs for privacy and security issues and uses Microsoft's Accessibility Insights Tool \cite{microsoft-2022} to analyze the accessibility of an app.

Study S8 analyses privacy and security-related concerns from user reviews of popular Android games. It manually analyses the reviews and labels them to create a training set for further analysis and then employs Convolutional Neural Networks (CNN) \cite{o2015introduction} with active learning for automated classification of reviews. Additionally, it analyses the relationship between app ratings and privacy-security-related reviews, suggesting that the current star rating system does not effectively capture users' concerns.

Study S9 addresses fairness, and accessibility concerns by analyzing tweets related to sharing economy platforms. It uses keyword search and manual analysis with the initial seed of {discr, bigot, prejud} to identify relevant tweets. Then it utilizes a Feature-Goal Interdependency Graph (F-SIG) \cite{noorian2014addressing} to design a conceptual crowd-driver model. It illustrates the relationship between user concerns, goals, and system features that are manually extracted from the tweets. This approach helps to visualize how different system features could potentially mitigate or exacerbate user concerns. Study S16 also addresses user concerns related to the sharing economy platform. It considers one and two-star rated user reviews of the Airbnb app and performs summarization of reviews using the LDA algorithm \cite{jelodar2019latent}.

Study S11 analyses user reviews of popular metaverse apps to uncover concerns related to security, safety, fairness, transparency, accessibility, and social solidarity. It applies topic modeling algorithm LDA \cite{jelodar2019latent} on the reviews with a star rating of three or less, to analyze and summarize the underlying themes.

Study S13 addresses privacy, transparency, fairness, and accessibility concerns by analyzing reviews of top health apps. It identifies and classifies reviews using the BoW \cite{zhang2010understanding} approach with Thesaurus API\footnote{\url{https://www.api-ninjas.com/api/thesaurus}}. It also included manual analysis for any wrong classification and updating the list of keywords accordingly to improve performance.

Study S14 analyzes security and privacy-related concerns by identifying reviews using keywords from permission documents. It manually labels the reviews to create the training data for automatically classifying reviews using the SVM \cite{yue2003svm} model with BoW \cite{zhang2010understanding} and character n-gram feature extraction. Additionally, it performs static app analysis to map the identified security and privacy-related reviews with subsequent security and privacy updates. This involves extracting permissions from app manifest files, scanning bytecode for APIs requiring dangerous permissions, and using tools like LibScout to check for third-party libraries. Then it tracks the changes in app versions to determine if privacy updates are triggered by the identified SPR reviews.

Studies S15 and S36 analyze user concerns related to fairness and transparency. Study S15 analyzes problematic microtransactions in gaming apps. It employs quantitative content analysis \cite{krippendorff2018content} to analyze one-star reviews of Android games and negative reviews of desktop games and identify problematic microtransactions. Additionally, the analysis includes both prevalence and salience scoring. Prevalence analysis determines how often a problematic microtransaction is discussed in more than two negative reviews across multiple games. Salience analysis measures the prominence of a problematic microtransaction by calculating the ratio of its occurrence across all reviews of a game to the total number of games where it appears. Study S36 conducts a large-scale analysis of AI-based apps. It uses keyword search and KeyBERT \cite{grootendorst2020keybert} with an initial seed of {fair, discrimination, bias} keywords to identify relevant reviews. It then manually labels the reviews and explores LR \cite{lavalley2008logistic}, SVM \cite{yue2003svm}, XGBoost (XGB) \cite{chen2015xgboost}, BERT \cite{devlin2018bert}, RoBERTa \cite{liu2019roberta}, and GPT-3 \cite{brown2020language} for automated classification. As a next step, it employs the K-means algorithm \cite{ahmed2020k} to summarize the concern-related reviews. Additionally, it manually analyzes the developer's replies for the top reviews to understand the common root causes for transparency and fairness concerns.

Study S18 analyzes accessibility concerns by relevant reviews using keyword search and manual analysis. It also employs the BERTopic \cite{grootendorst2022bertopic} to summarize the accessibility-related concerns. Study S44 builds upon this to further analyze the accessibility-related reviews to understand the impact of updates on the accessibility of the app. It uses keyword search to find the reviews related to app updates and employs ChatGPT-4\footnote{\url{https://openai.com/product/gpt-4}} to analyze the data using specific prompts.

Studies S39, S38, S37, and S42 analyze human values violations from user reviews. Study S42 also considers GitHub to understand the alignment of the developers' discussions with the user values. Study S39 uses Schwartz’s values theory \cite{schwartz1992universals} to curate a list of keywords and identify relevant user reviews. Studies S38 and S37 conduct further research and use keyword search to identify and manually label the reviews with honesty violations. Both the studies explore LR \cite{lavalley2008logistic}, SVM \cite{yue2003svm}, DT \cite{kingsford2008decision}, GBT \cite{friedman2001greedy}, Neural Network (NN) \cite{abdi1999neural}, Deep Neural Network {DNN} \cite{nielsen2015neural}, and Generative Adversarial Network (GAN) \cite{goodfellow2014generative} for automated classification of reviews. Additionally, study S37 conducts developer surveys and interviews to identify the potential causes of violations, their consequences, and strategies to mitigate them. Study S42 compares the reviews of open-source apps with the developer discussions on GitHub by using manual analysis and keyword search to identify relevant reviews and discussions. It also trains models such as LR \cite{lavalley2008logistic}, SVM \cite{yue2003svm}, RF \cite{rigatti2017random}, XGB \cite{chen2015xgboost}, BERT \cite{devlin2018bert}, RoBERTa \cite{liu2019roberta}, and DistilBERT \cite{sanh2019distilbert} with different combinations of feature extraction algorithms TF-IDF \cite{yun2005improved}, and Word2Vec \cite{church2017word2vec} for automated classification.

Study S23 performs a manual analysis of reviews of exam-proctoring browser extensions to analyze the security and privacy concerns of students. Additionally, it surveys exam-taking students to understand their awareness and exposure to the proctoring services.

Studies S19, S25, S28, and S46 analyze accessibility-related concerns from user reviews of Android apps. Study S25 uses a keyword search to identify relevant reviews and then performs manual analysis to understand the diversity of accessibility reviews. Study 19 furthers this research to automatically identify accessibility reviews. It uses FeatureHashing \cite{shi2009hash} for feature extraction and explores LR \cite{lavalley2008logistic}, Decision Forest (DF) \cite{prinzie2008random}, Boosted Decision Trees (BDTs) \cite{friedman2001greedy}, NN \cite{abdi1999neural}, SVM \cite{yue2003svm}, Averaged Perceptron (AP) \cite{collins2002discriminative}, Bayes Point Machine (BPM) \cite{herbrich2001bayes}, Decision Jungle (DJ) \cite{shotton2013decision}, and Locally Deep-SVM (LDSVM) \cite{jose2013local} algorithms for automated classification. Study S46 also furthers study S25 to identify different accessibility categories using deductive coding \cite{fereday2006demonstrating}. Study S28 analyzes user reviews of COVID-19 apps by using keyword search and manual analysis to identify relevant reviews. It also considers app description, version history logs, and other details submitted by app developers and performs manual analysis to evaluate and recommend accessibility-related features.

Study S26 analyzes user reviews to identify violations of app stores' market policies. First, it manually identifies twenty-six unique topics from market policies, curates the list of keywords and subsequently manually labels the relevant reviews with these identified topics. It then employs Biterm Topic Modeling (BTM) and semantic rule-based classification to automatically classify the reviews into predefined themes.

Studies S29 and citS32 analyze a range of concerns related to ethical requirements from user reviews. Study S29 addresses the concerns of marginalized communities by analyzing Reddit posts. It uses keyword search to identify posts related to any social platform and then performs manual analysis to label the posts with the related concern. It then explores models such as LR \cite{lavalley2008logistic}, SVM \cite{yue2003svm}, Linear Support Vector Classification (LinearSVC) \cite{alpaydin2020introduction}, NuSVC \cite{alpaydin2020introduction}, and Gaussian NB (GNB) \cite{alpaydin2020introduction} to automate the classification of posts into the concern category. Study S32 considers reviews of Android apps to analyze user concerns. It performs manual labeling of reviews into specific categories of concern and explores RF \cite{rigatti2017random}, Multinomial NB (MNB) \cite{losada2008assessing}, LR \cite{lavalley2008logistic}, Multilayer Perceptron (MLP) \cite{popescu2009multilayer}, SVM \cite{yue2003svm}, and BERT \cite{devlin2018bert} for automated classification.

Study S33 conducts large-scale analysis of user reviews to address privacy, security, transparency, fairness, and social solidarity concerns. It uses random sampling to identify and label relevant reviews and employs pre-trained large language models GPT-2 \cite{radford2023robust}, BERT \cite{devlin2018bert}, RoBERTa \cite{liu2019roberta}, XLM-RoBERTa \cite{conneau2019unsupervised}, and BART \cite{lewis2019bart} for automated classification.

Study S34 addresses the social impact of generative AI by analyzing the user reviews of graphic designing tools. It uses BERTopic and manual analysis to identify the relevant reviews and then performs thematic analysis \cite{braun2012thematic} to identify the reviews based on different use cases of tools that have a societal impact.

Study S41 analyzes gender-based reviews to address related concerns. It uses keyword search with KeyBERT \cite{grootendorst2020keybert} and manual analysis to identify the relevant reviews and label them with the defined categories. It explores LR \cite{lavalley2008logistic}, SVM-LR \cite{cortes1995support}, RF \cite{rigatti2017random}, XGB \cite{chen2015xgboost}, AdaBoost \cite{freund1997decision}, DT \cite{kingsford2008decision}, CNN \cite{o2015introduction}, LSTM \cite{hochreiter1997long}, and RoBERTa \cite{liu2019roberta} models for automated classification of reviews. It then performs the manual analysis of identified reviews to understand the underlying issues within the gender-based reviews.

\begin{table*}[]
    \centering
    \caption{List of primary studies and the concerns they discuss. The categorization of studies into ethical requirements is done based on the concern they address.}
    \label{tab:concerns}
    \begin{tabular}{|p{1.8cm}|c|c|c|c|c|c|c|c|}
            \hline
            \textbf{Reference} & \textbf{Privacy} & \textbf{Security} & \textbf{Accessibility} & \textbf{Accountability} & \textbf{Transparency} & \textbf{Fairness} & \textbf{Safety} & \textbf{Social Solidarity} \\
            \hline
            \hline
            S1 & \faIcon[solid]{square} & \faIcon[]{square} & \faIcon[solid]{square} & \faIcon[solid]{square} & \faIcon[solid]{square} & \faIcon[solid]{square} & \faIcon[solid]{square} & \faIcon[solid]{square} \\
            \hline
            S2 & \faIcon[]{square} & \faIcon[]{square} & \faIcon[]{square} & \faIcon[]{square} & \faIcon[]{square} & \faIcon[]{square} & \faIcon[solid]{square} & \faIcon[]{square} \\
            \hline
            S3 S4 S8 S14 S20 S21 S22 S23 S24 S31 S35 S40 S43 & \faIcon[solid]{square} & \faIcon[solid]{square} & \faIcon[]{square} & \faIcon[]{square} & \faIcon[]{square} & \faIcon[]{square} & \faIcon[]{square} & \faIcon[]{square} \\
            \hline
            S5 S17 S27 S45 S47 & \faIcon[solid]{square} & \faIcon[]{square} & \faIcon[]{square} & \faIcon[]{square} & \faIcon[]{square} & \faIcon[]{square} & \faIcon[]{square} & \faIcon[]{square} \\
            \hline
            S6 & \faIcon[solid]{square} & \faIcon[]{square} & \faIcon[]{square} & \faIcon[solid]{square} & \faIcon[solid]{square} & \faIcon[solid]{square} & \faIcon[solid]{square} & \faIcon[]{square} \\
            \hline
            S7 & \faIcon[solid]{square} & \faIcon[solid]{square} & \faIcon[solid]{square} & \faIcon[solid]{square} & \faIcon[]{square} & \faIcon[]{square} & \faIcon[]{square} & \faIcon[]{square} \\
            \hline
            S9 & \faIcon[]{square} & \faIcon[]{square} & \faIcon[solid]{square} & \faIcon[]{square} & \faIcon[]{square} & \faIcon[solid]{square} & \faIcon[]{square} & \faIcon[]{square} \\
            \hline
            S10 & \faIcon[solid]{square} & \faIcon[]{square} & \faIcon[]{square} & \faIcon[]{square} & \faIcon[solid]{square} & \faIcon[]{square} & \faIcon[]{square} & \faIcon[solid]{square} \\
            \hline
            S11 & \faIcon[]{square} & \faIcon[solid]{square} & \faIcon[]{square} & \faIcon[solid]{square} & \faIcon[solid]{square} & \faIcon[solid]{square} & \faIcon[solid]{square} & \faIcon[solid]{square} \\
            \hline
            S12 S15 S36 & \faIcon[]{square} & \faIcon[]{square} & \faIcon[]{square} & \faIcon[]{square} & \faIcon[solid]{square} & \faIcon[solid]{square} & \faIcon[]{square} & \faIcon[]{square} \\
            \hline
            S13 & \faIcon[solid]{square} & \faIcon[]{square} & \faIcon[solid]{square} & \faIcon[]{square} & \faIcon[solid]{square} & \faIcon[solid]{square} & \faIcon[]{square} & \faIcon[]{square} \\
            \hline
            S16 & \faIcon[]{square} & \faIcon[]{square} & \faIcon[]{square} & \faIcon[solid]{square} & \faIcon[]{square} & \faIcon[]{square} & \faIcon[solid]{square} & \faIcon[]{square} \\
            \hline
            S18 S44 & \faIcon[]{square} & \faIcon[]{square} & \faIcon[solid]{square} & \faIcon[]{square} & \faIcon[]{square} & \faIcon[]{square} & \faIcon[]{square} & \faIcon[]{square} \\
            \hline
            S19 S25 S28 S46 & \faIcon[]{square} & \faIcon[]{square} & \faIcon[solid]{square} & \faIcon[]{square} & \faIcon[]{square} & \faIcon[]{square} & \faIcon[]{square} & \faIcon[]{square} \\
            \hline
            S26 & \faIcon[solid]{square} & \faIcon[solid]{square} & \faIcon[]{square} & \faIcon[]{square} & \faIcon[]{square} & \faIcon[]{square} & \faIcon[solid]{square} & \faIcon[]{square} \\
            \hline
            S29 & \faIcon[solid]{square} & \faIcon[]{square} & \faIcon[]{square} & \faIcon[]{square} & \faIcon[]{square} & \faIcon[solid]{square} & \faIcon[solid]{square} & \faIcon[solid]{square} \\
            \hline
            S30 & \faIcon[solid]{square} & \faIcon[solid]{square} & \faIcon[]{square} & \faIcon[solid]{square} & \faIcon[]{square} & \faIcon[]{square} & \faIcon[solid]{square} & \faIcon[]{square} \\
            \hline
            S32 & \faIcon[solid]{square} & \faIcon[solid]{square} & \faIcon[solid]{square} & \faIcon[solid]{square} & \faIcon[solid]{square} & \faIcon[solid]{square} & \faIcon[solid]{square} & \faIcon[solid]{square} \\
            \hline
            S33 & \faIcon[solid]{square} & \faIcon[solid]{square} & \faIcon[]{square} & \faIcon[solid]{square} & \faIcon[solid]{square} & \faIcon[solid]{square} & \faIcon[]{square} & \faIcon[solid]{square} \\
            \hline
            S34 & \faIcon[]{square} & \faIcon[]{square} & \faIcon[]{square} & \faIcon[]{square} & \faIcon[]{square} & \faIcon[]{square} & \faIcon[]{square} & \faIcon[solid]{square} \\
            \hline
            S37 S38 & \faIcon[]{square} & \faIcon[]{square} & \faIcon[]{square} & \faIcon[solid]{square} & \faIcon[solid]{square} & \faIcon[solid]{square} & \faIcon[]{square} & \faIcon[]{square} \\
            \hline
            S39 & \faIcon[solid]{square} & \faIcon[solid]{square} & \faIcon[]{square} & \faIcon[]{square} & \faIcon[]{square} & \faIcon[]{square} & \faIcon[]{square} & \faIcon[solid]{square} \\
            \hline
            S41 & \faIcon[]{square} & \faIcon[]{square} & \faIcon[]{square} & \faIcon[]{square} & \faIcon[]{square} & \faIcon[solid]{square} & \faIcon[solid]{square} & \faIcon[]{square} \\
            \hline
            S42 & \faIcon[solid]{square} & \faIcon[solid]{square} & \faIcon[solid]{square} & \faIcon[]{square} & \faIcon[]{square} & \faIcon[]{square} & \faIcon[]{square} & \faIcon[]{square} \\
            \hline
    \end{tabular}
\end{table*}

\begin{tcolorbox}[arc=0mm,width=\columnwidth,
    top=1mm,left=1mm,  right=1mm, bottom=1mm,
    boxrule=1pt] 
Summary of RQ3: Keyword-based approach has been identified as the most commonly used technique to extract relevant user reviews. The two major downstream tasks that have been explored the most are the classification of reviews and the summarization of reviews. Due to the advancements in ML and AI, various state-of-the-art techniques have been utilized for these tasks.
\end{tcolorbox}

\subsection{\textbf{RQ4: Implications to address user concerns}}
Here we present the implications, guidelines, or suggestions from primary studies that can be helpful to address the user concerns. We have presented these details based on the ethical requirements.

\textit{\underline{Privacy}}: Privacy requirement determines the minimum amount of personal data to be collected, its purpose and informed consent for the collection, its limited usage, and individual's access to their data \cite{wright2011framework}. 

Study S5 suggests that software development organizations should be aware of narratives that could shape user perceptions and privacy concerns. It demonstrates that user behavior and their concerns could be influenced by narratives, which may not always be an accurate reflection of reality. Hence, an organization should allocate resources
in considering these macro-level narratives that could greatly impact an organization’s software. Since a narrative represents an aggregate of user feedback, an organization can develop requirements from a narrative similar to how it may develop requirements from several user feedbacks at a time. It illustrates that an organization may expand its requirements elicitation process to include the addition of user discussion data from Reddit and analyzing it for privacy narratives. The added insights from the requirements elicitation may facilitate improvements to the organization’s understanding of user
privacy concerns and help it develop more relevant privacy requirements. 

\par Study S17 suggests that organizations should design tools that provide a clear understanding of data-collection practices, similar to Apple Privacy Nutrition Labels \cite{apple-inc-2020}. Studies S21, S22, S27, and S32 suggest that software developers should only request permissions that are pivotal to performing operations in the app and should carefully clarify the needs for it. Study S23 suggests that institutions and educators should follow a principle of least monitoring by using the minimum number of monitoring types necessary in online exam proctoring software. Study S42 suggests developing privacy-relevant test cases from reviews to support app testing.

\textit{\underline{Security}}: Security requirement determines the measures to be taken to ensure the protection of personal data, for example utilizing encryption and/or access control \cite{wright2011framework}.

Study S7 suggests using tools such as RiskInDroid \cite{merlo2017riskindroid}, AOSP \cite{aafer2018precise}, DYNAMO \cite{dawoud2021bringing}, Stowaway \cite{felt2011android}, PScout \cite{au2012pscout}, and Droidtector \cite{wu2019overprivileged} to analyze app permissions and adhere to best practices to ensure security, compliance, and user trust while meeting regulatory guidelines. Additionally, it suggests users take precautionary actions like following Android best practices for security \cite{oh2012best}, to protect their data. Study S23 suggests that institutions should thoroughly review common vulnerabilities and exposures of the online exam proctoring software they plan to license for installation on students’ personal computers.

\textit{\underline{Accessibility}}: Accessibility requirement determines the inclusivity and user-friendliness of software for all citizens, including senior citizens and people with disabilities \cite{wright2011framework}.

Study S6 suggests that developers of mental health apps should develop a simple interactive checklist interface that can be used to take user symptoms and other preferences as input and recommendation algorithms can be adopted to show them an ordered list of therapists based on their symptoms. Study S25 suggests that software developers should allow users to customize several aspects of the user interface of the software: color modes, text, and widget sizes. Study S28 proposes a tool named Emerging Apps Accessibility Evaluator and Recommender (EAAER), which can be used as a mandatory requirement for developers to evaluate the accessibility of emerging apps before publishing them to any app stores. Study S30 suggests that developers should develop a user interface such that it is simple, easy to use and navigate, and customizable. Study 19 calls for innovative methods that can support technology professionals in maintaining the accessibility of their app after its release.

\textit{\underline{Accountability}}: Accountability requirement determines the joint responsibilities of all stakeholders at different stages of development to address both positive and negative outcomes of software \cite{wright2011framework}.

Study S16 suggests that it is paramount to provide sharing economy users with customer support and a service/resolution center to help resolve any issues that may occur. Study S30 suggests that developers should provide diverse user support mechanisms, such as an instant messaging feature (to connect users with customer service agents), an in-app contact form, and an easy-to-understand tutorial/help feature.

\textit{\underline{Transparency}}: Transparency requirement determines the level of openness and disclosure necessary to build public trust and confidence in software \cite{wright2011framework}. 

Study S6 suggests that developers should carefully design the interfaces that make pricing information transparent. For instance, they can adopt simple visualization approaches that show the pricing breakdown and total cost upfront and provide reminders and warning notifications before any type of payment is charged. Study S10 suggests following these ethical guidelines for developing freemium mental health apps: (a) clarity about the exact medical condition being addressed, (e.g. the app targets depression, but not anxiety or bipolar); (b) precise descriptions of overall treatment features (e.g. provides access to volunteer support but not trained therapists), (c) specific details of charging practices (e.g. length of time for which features are free, and whether users have to actively unsubscribe when they no longer need the app) and (d) whether apps include ads which may potentially impair treatments. 

\textit{\underline{Fairness}}: Fairness requirement determines the anti-discriminatory measures to prevent harm or disadvantage to individuals, such as discriminatory practices in pricing or service access \cite{wright2011framework}.

Study S9 suggests a domain model that represents the complex relation among various requirements of the sharing economy platforms to address the user concerns at the RE phase. Studies S12 and S30 suggest developing more flexible and fair payment models for mental health apps that account for unexpected circumstances that may affect the provision of care and that can cater to everyone, especially young adults and people with financial challenges. Study S29 suggests devoting more resources to equitable and objective informational resources. Study S33 suggests that developers should avoid completely relying on AI for decision-making, whether it is adhering to community standards or generating recommendations, because of its unfair results. Study S41 suggests that developers should develop mechanisms to empower users to apply their gender-related preferences and expectations in apps in a meaningful way.

\textit{\underline{Safety}}: Safety requirement determines the measures to be taken to protect individuals from physical, psychological, and economic harm \cite{wright2011framework}.

Study S29 suggests adding stricter regulations for advertisers and harsher rules regarding scams by independent creators on social platforms. Additionally, it suggests that developers should change how they deal with nonconsensual pornography on social platforms to remove the burden from the victims and put it back on the perpetrators. Study S32 suggests that developers should involve other departments (legal, public relations) while crafting a safety solution where there is a potential risk for serious harm to the user, litigation, and damage to the platform’s reputation. Additionally, it suggests that software professionals should contact proper authorities or otherwise quickly intervene to ensure the safety of their users (consider situations regarding sexual assault, self-harm, or other acts of physical violence).

\textit{\underline{Social Solidarity}}: Social solidarity requirement determines the wider social good of software. It ensures that all the stakeholders get an equal opportunity to voice their concerns, individuals are not manipulated in any sense for using the software, the software does not cause any psychological harm, it does not lead to stigmatization or be seen as a substitute for human contact, there impacting the social ties \cite{wright2011framework}.

Study S1 suggests that it is not enough to solely focus on the app itself, but rather developers must consider the interrelated elements around apps for depression that contribute to user experiences and ethical implications. It captures the key elements for the ethical design of apps for depression that focus on overall trust and social good. Alternatively, it favors the responsible innovation approach \cite{owen2013framework} which encourages designers to use moral conflicts to inspire, rather than hinder innovation \cite{van2013value}. Study S6 suggests employing community support using peer-led moderation in mobile mental health apps and keeping them accountable for any negative outcomes. Study S29 suggests creating design features that mimic real-life social interaction more closely to help users build healthier relationships on social platforms. Study S33 suggests that developers should consider the cultural background of users to make it more inclusive, because they may prefer software features that accommodate their situation. Study S39 suggests that a critical technical practice would encourage positive interaction between the social, cultural, and technical aspects of software. This would involve taking a step outside the technical field of specialization and would limit the challenges of trying to apply technical mindsets to non-technical problems \cite{agre2014toward}. Additionally, it will entail engaging with the social and human sciences approach such as critical theory \cite{agre2014toward} to open up the assumptions underpinning SE and how they affect the social and cultural aspects of society, and in so doing support the evaluation of the SE field’s contribution to society.

\begin{tcolorbox}[arc=0mm,width=\columnwidth,
    top=1mm,left=1mm,  right=1mm, bottom=1mm,
    boxrule=1pt] 
Summary of RQ4: Software development should encompass a wide range of ethical considerations to address user concerns and promote responsible innovation. These considerations span privacy, security, accessibility, accountability, transparency, fairness, safety, and social solidarity. 
\\
\par Software developers are encouraged to minimize data collection, use security analysis tools, create customizable and accessible interfaces, provide robust customer support, offer clear pricing and implement fair payment models, protect users from harm, and consider the broader social and cultural impacts of their software. By integrating these requirements into the development process, developers can build trustworthy, inclusive, and socially beneficial software that better serves users' needs while mitigating potential concerns.
\end{tcolorbox}

\section{Discussion} \label{discuss}
The software ecosystem and user reviews are rich sources of information regarding user experience and expectations. By utilizing this data, developers can enhance their understanding of their target audience. Extracting insights from user reviews can be beneficial for developers in gathering feedback on the ethical considerations of their software. However, analyzing ethical concerns within user reviews necessitates pre-processing at the content level, such as filtering out irrelevant information and assessing the credibility and authenticity of the reviews and their sources. While there is currently a limited number of research studies that scrutinize user reviews from an ethical perspective, the emerging trends and findings are encouraging. Therefore, it is anticipated that with further investigation, user reviews will offer a more precise insight into the ethical concerns of software from the RE perspective. Here, we present the principal findings from the SLR, knowledge gaps, and open challenges in this domain.

\subsection{\textbf{Principal findings}}
The findings of the SLR shed light on the techniques used to derive user concerns related to the ethical requirements of the software, the implications or suggestions for developers to address these concerns, and the types of datasets used in this regard.

- Most of the studies are large-scale and do not focus on specific software categories. However, analyzing a specific software category could help to understand the complex ethical landscape from the RE perspective as demonstrated by study S9.

- Most studies have used the Google Play store as a user review data source but integrating reviews from other data sources could provide more insights because Google Play has a limitation on the length of the review content as demonstrated by study S29.

- Privacy and security were the most studied ethical concerns among all and very few studies have analyzed a diverse set of ethical concerns from user reviews as shown in Table~\ref{tab:concerns}. Most of the studies were exploratory, based on manual analysis and automated classification. Most of the studies have used a keyword search approach to identify the relevant reviews but this approach has limitations as it can identify irrelevant reviews also, as demonstrated by study S47.

- Most of the studies have presented automated classification and either manual or automated analysis of reviews. This helps in identifying and understanding the user concerns in brief but it doesn't help to translate it to requirements that can be implemented at the RE phase to address these concerns. Most of the studies have presented suggestions that can be considered by developers while designing software but only one study S9 has presented a model to address user concerns at the RE phase.

- Few of the studies have open-sourced their datasets which can be used by the community for further research. The links to the data sources are presented in Appendix \ref{appendixD}.

- Examples of the user reviews, from primary studies, mentioning concerns related to ethical requirements are given in Appendix \ref{appendixD}.

\subsection{\textbf{Knowledge gaps}}
Two major knowledge gaps are identified from this SLR: limitations with keyword-based search for identifying concern-related reviews and lack of RE perspective from user reviews analysis. 

1) To detect user reviews related to concerns, the majority of the studies have utilized a keyword-based search method, in which keywords are connected to the ethical concerns being discussed. Some studies have also utilized manual evaluation of user reviews. However, both of these methods have their limitations. Keyword-based searches may not be applicable across different contexts and may result in a high number of false positives, while manual analysis is impractical for analyzing a large volume of user reviews. Therefore, there is a need for a more generalizable approach that can effectively identify user reviews associated with various concerns and minimize the need for manual analysis.

2) There is limited research on translating user ethical concerns into concrete software requirements for developers. While automation tools can identify concern-related reviews, developers may not have the time to read and comprehend all the reviews to extract actionable requirements. To effectively address user concerns during the RE phase, these concerns need to be translated into clear, understandable requirements. Only one study (S9) discusses this translation process, highlighting the need for further research in this area to ensure ethical concerns are adequately addressed in the SDLC, specifically the RE phase.

\subsection{\textbf{Open challenges}}
Two major challenges exist to bridge the identified knowledge gaps.

- Very few user reviews explicitly report ethical concerns. Determining which reviews are pertinent to ethical concerns is inherently difficult due to how users express these issues. Additionally, a single user review can report multiple concerns, complicating the extraction and categorization process. So, identifying relevant reviews and categorizing them into related ethical concerns remains the first biggest challenge in this subject.

- The second major challenge remains with translating identified ethical concerns into software requirements, because of the unstructured nature of the user reviews and the colloquial language used.

\section{Threats to validity} \label{limit}
Following the guidelines for conducting SLR presented in \cite{keele2007guidelines}, in the following the threats to validity that may have affected the results presented in this SLR are analyzed. The categories of threats to validity defined by Zhou et al. \cite{7890583} have been analyzed:

\subsection{Construct validity}
Construct validity refers mainly to the review protocol and the degree of inclusion of the selected search string, digital libraries, and selection criteria concerning the research questions. To minimize the threats to construct validity, a validation procedure was included for each step of the review protocol. For instance, the search string was defined as generic as possible not to exclude any potentially interesting studies following PICO guidelines. To mitigate potential bias from our choice of libraries and ensure no studies were overlooked, we employed one-step backward and forward snowball sampling. Despite our efforts, the possibility remains that some key studies might have been overlooked.

\subsection{Internal Validity}
Internal validity refers to the reliability of the cause-effect relationship between the results and the methodology of the review. In this SLR, the search procedure was mainly performed by a single researcher, so the results may be subject to selection bias. To reduce this, a validation procedure was performed in which the second author reviewed a randomly selected set of articles so that each study was 
reviewed by both the authors.

\subsection{External Validity}
External validity refers to how the results of the review apply to the subject area, which in this case is addressing the user concerns related to ethical requirements. We derived eight ethical requirements from the literature and categorized the primary studies into these requirements based on how they addressed user ethical concerns. Although this categorization was carefully done, it is inherently subjective and dependent on our interpretation of the studies' content. Additionally, the dynamic and evolving nature of ethical considerations in SE means that new ethical concerns could emerge that were not covered by these predefined categories, potentially impacting the generalizability of this SLR.

\subsection{Conclusion validity}
Conclusion validity refers to the reproducibility of the results. To achieve this goal, we defined clear search strings, steps, and procedures; and we provided an explicit list of clear inclusion and exclusion criteria. We performed each step automatically wherever possible. However, the subjectivity of conducting parts of the review prevents us from guaranteeing obtaining the exact same results from performing the review by other researchers, we believe that the overall results should be similar.

\section{Conclusion} \label{conclude}

In this study, we elaborated on the need and importance of ethical software and reviewed the different requirements to make software ethical and trustworthy. We then performed an SLR addressing user concerns about the ethical requirements, different techniques to analyze them, and their corresponding implications. By analyzing 47 primary studies, we identified a list of techniques commonly used in identifying and analyzing user reviews related to concerns and a set of guidelines that can help address user concerns. In addition, the primary studies were classified into ethical requirements based on the corresponding concerns they analyzed from user reviews.

Our SLR provided a deeper understanding of current challenges in ethical requirements and extraction is user reviews. Based on the findings, we propose potential directions for further research which aim to stimulate reflection among researchers and professionals on the importance of incorporating user reviews in the RE phase to address ethical concerns and make software ethical and trustworthy.

\bibliographystyle{IEEEtran}
\bibliography{references}

\newpage
\onecolumn

\begin{appendices}

\section{Selected studies} \label{appendixA}

\begin{longtblr}[
    caption={List of primary studies},
    label={tab:ps}
]{
  colspec = {lp{14cm}},
  rowhead = 1,
}
    \hline
    \textbf{Study} & \textbf{Reference} \\
    \hline
    S1 & Bowie-DaBreo, Dionne, Corina Sas, Heather Iles-Smith, and Sandra Sünram-Lea. ``User perspectives and ethical experiences of apps for depression: a qualitative analysis of user reviews." In Proceedings of the 2022 CHI Conference on Human Factors in Computing Systems, pp. 1-24. 2022. \\
    S2 & O'Hagan, Joseph, Florian Mathis, and Mark McGill. ``User Reviews as a Reporting Mechanism for Emergent Issues Within Social VR Communities." In Proceedings of the 22nd International Conference on Mobile and Ubiquitous Multimedia, pp. 236-243. 2023. \\
    S3 & Kong, Deguang, Lei Cen, and Hongxia Jin. ``Autoreb: Automatically understanding the review-to-behavior fidelity in android applications." In Proceedings of the 22nd ACM SIGSAC conference on computer and communications security, pp. 530-541. 2015. \\
    S4 & Ebrahimi, Fahimeh, and Anas Mahmoud. ``Unsupervised summarization of privacy concerns in mobile application reviews." In Proceedings of the 37th IEEE/ACM International Conference on Automated Software Engineering, pp. 1-12. 2022. \\
    S5 & Li, Ze Shi, Manish Sihag, Nowshin Nawar Arony, Joao Bezerra Junior, Thanh Phan, Neil Ernst, and Daniela Damian. ``Narratives: the unforeseen influencer of privacy concerns." In 2022 IEEE 30th International Requirements Engineering Conference (RE), pp. 127-139. IEEE, 2022.\\
    S6 & Haque, Md Romael, and Sabirat Rubya. ```` For an app supposed to make its users feel better, it sure is a joke"-an analysis of user reviews of mobile mental health applications." Proceedings of the ACM on Human-Computer Interaction 6, no. CSCW2 (2022): 1-29.\\
    S7 & Kishnani, Urvashi, Naheem Noah, Sanchari Das, and Rinku Dewri. ``Assessing Security, Privacy, User Interaction, and Accessibility Features in Popular E-Payment Applications." In Proceedings of the 2023 European Symposium on Usable Security, pp. 143-157. 2023.\\
    S8 & Lee, Sangho, Daekwon Pi, Junho Jang, and Huy Kang Kim. ``Poster: What Can Review with Security Concern Tell Us before Installing Apps?." In Adjunct Proceedings of the 2022 ACM International Joint Conference on Pervasive and Ubiquitous Computing and the 2022 ACM International Symposium on Wearable Computers, pp. 70-71. 2022.\\
    S9 & Tushev, Miroslav, Fahimeh Ebrahimi, and Anas Mahmoud. ``Digital discrimination in sharing economy a requirements engineering perspective." In 2020 IEEE 28th International Requirements Engineering Conference (RE), pp. 204-214. IEEE, 2020.\\
    S10 & Eagle, Tessa, Aman Mehrotra, Aayush Sharma, Alex Zuniga, and Steve Whittaker. ```` Money doesn't buy you happiness": negative consequences of using the freemium model for mental health apps." Proceedings of the ACM on Human-Computer Interaction 6, no. CSCW2 (2022): 1-38.\\
    S11 & Çallı, Büşra Alma, and Çağla Ediz. ``Top concerns of user experiences in Metaverse games: A text-mining based approach." Entertainment Computing 46 (2023): 100576.\\
    S12 & Jo, Eunkyung, Whitney-Jocelyn Kouaho, Stephen M. Schueller, and Daniel A. Epstein. "Exploring User Perspectives of and Ethical Experiences With Teletherapy Apps: Qualitative Analysis of User Reviews." JMIR Mental Health 10 (2023): e49684.\\
    S13 & Haggag, Omar, John Grundy, Mohamed Abdelrazek, and Sherif Haggag. ``A large scale analysis of mHealth app user reviews." Empirical Software Engineering 27, no. 7 (2022): 196.\\
    S14 & Nguyen, Duc Cuong, Erik Derr, Michael Backes, and Sven Bugiel. "Short text, large effect: Measuring the impact of user reviews on android app security \& privacy." In 2019 IEEE symposium on Security and Privacy (SP), pp. 555-569. IEEE, 2019.\\
    S15 & Petrovskaya, Elena, Sebastian Deterding, and David I. Zendle. ``Prevalence and salience of problematic microtransactions in top-grossing mobile and PC games: a content analysis of user reviews." In Proceedings of the 2022 CHI Conference on Human Factors in Computing Systems, pp. 1-12. 2022.\\
    S16 & Al-Ramahi, Mohammad, and Ali Ahmed. ``Identifying Users' Concerns in Lodging Sharing Economy Using Unsupervised Machine Learning Approach." In 2019 2nd International Conference on Data Intelligence and Security (ICDIS), pp. 160-166. IEEE, 2019.\\
    S17 & Nema, Preksha, Pauline Anthonysamy, Nina Taft, and Sai Teja Peddinti. ``Analyzing user perspectives on mobile app privacy at scale." In Proceedings of the 44th International Conference on Software Engineering, pp. 112-124. 2022.\\
    S18 & Oliveira, Alberto Dumont Alves, Paulo Sérgio Henrique Dos Santos, Wilson Estécio Marcílio Júnior, Wajdi M. Aljedaani, Danilo Medeiros Eler, and Marcelo Medeiros Eler. ``Analyzing Accessibility Reviews Associated with Visual Disabilities or Eye Conditions." In Proceedings of the 2023 CHI Conference on Human Factors in Computing Systems, pp. 1-14. 2023.\\
    S19 & AlOmar, Eman Abdullah, Wajdi Aljedaani, Murtaza Tamjeed, Mohamed Wiem Mkaouer, and Yasmine N. El-Glaly. ``Finding the needle in a haystack: On the automatic identification of accessibility user reviews." In Proceedings of the 2021 CHI conference on human factors in computing systems, pp. 1-15. 2021.\\
    S20 & Hatamian, Majid, and Jetzabel Serna. ``ARM: ANN-based ranking model for privacy and security analysis in smartphone ecosystems." In 2017 International Carnahan Conference on Security Technology (ICCST), pp. 1-6. IEEE, 2017.\\
    S21 & Hatamian, Majid, Jetzabel Serna, and Kai Rannenberg. ``Revealing the unrevealed: Mining smartphone users privacy perception on app markets." Computers \& Security 83 (2019): 332-353.\\
    S22 & Kishnani, Urvashi, Naheem Noah, Sanchari Das, and Rinku Dewri. ``Privacy and security evaluation of mobile payment applications through user-generated reviews." In Proceedings of the 21st Workshop on Privacy in the Electronic Society, pp. 159-173. 2022.\\
    S23 & Balash, David G., Dongkun Kim, Darika Shaibekova, Rahel A. Fainchtein, Micah Sherr, and Adam J. Aviv. ``Examining the examiners: Students' privacy and security perceptions of online proctoring services." In Seventeenth symposium on usable privacy and security (SOUPS 2021), pp. 633-652. 2021.\\
    S24 & Tao, Chuanqi, Hongjing Guo, and Zhiqiu Huang. ``Identifying security issues for mobile applications based on user review summarization." Information and Software Technology 122 (2020): 106290.\\
    S25 & Eler, Marcelo Medeiros, Leandro Orlandin, and Alberto Dumont Alves Oliveira. ``Do Android app users care about accessibility? an analysis of user reviews on the Google play store." In Proceedings of the 18th Brazilian symposium on human factors in computing systems, pp. 1-11. 2019.\\
    S26 & Hu, Yangyu, Haoyu Wang, Tiantong Ji, Xusheng Xiao, Xiapu Luo, Peng Gao, and Yao Guo. ``Champ: Characterizing undesired app behaviors from user comments based on market policies." In 2021 IEEE/ACM 43rd International Conference on Software Engineering (ICSE), pp. 933-945. IEEE, 2021.\\
    S27 & Scoccia, Gian Luca, Stefano Ruberto, Ivano Malavolta, Marco Autili, and Paola Inverardi. ``An investigation into Android run-time permissions from the end users' perspective." In Proceedings of the 5th international conference on mobile software engineering and systems, pp. 45-55. 2018.\\
    S28 & Haggag, Omar, John Grundy, Mohamed Abdelrazek, and Sherif Haggag. ``Better Addressing diverse accessibility issues in emerging apps: A case study using covid-19 apps." In Proceedings of the 9th IEEE/ACM International Conference on Mobile Software Engineering and Systems, pp. 50-61. 2022.\\
    S29 & Olson, Lauren, Emitzá Guzmán, and Florian Kunneman. ``Along the margins: Marginalized communities’ ethical concerns about social platforms." In 2023 IEEE/ACM 45th International Conference on Software Engineering: Software Engineering in Society (ICSE-SEIS), pp. 71-82. IEEE, 2023.\\
    S30 & Oyebode, Oladapo, Felwah Alqahtani, and Rita Orji. ``Using machine learning and thematic analysis methods to evaluate mental health apps based on user reviews." IEEE Access 8 (2020): 111141-111158.\\
    S31 & Hatamian, Majid, Nurul Momen, Lothar Fritsch, and Kai Rannenberg. ``A multilateral privacy impact analysis method for android apps." In Privacy Technologies and Policy: 7th Annual Privacy Forum, APF 2019, Rome, Italy, June 13–14, 2019, Proceedings 7, pp. 87-106. Springer International Publishing, 2019.\\
    S32 & Olson, Lauren, Neelam Tjikhoeri, and Emitzá Guzmán. ``The Best Ends for the Best Means: Ethical Concerns in App Reviews." arXiv preprint arXiv:2401.11063 (2024).\\
    S33 & Arony, Nowshin Nawar, Ze Shi Li, Bowen Xu, and Daniela Damian. ``Inclusiveness Matters: A Large-Scale Analysis of User Feedback." arXiv preprint arXiv:2311.00984 (2023).\\
    S34 & Swift, Ian P., and Debaleena Chattopadhyay. ``A Value-Oriented Investigation of Photoshop's Generative Fill." arXiv preprint arXiv:2404.17781 (2024).\\
    S35 & Zhang, Jianzhang, Jinping Hua, Yiyang Chen, Nan Niu, and Chuang Liu. ``Mining User Privacy Concern Topics from App Reviews." arXiv preprint arXiv:2212.09289 (2022).\\
    S36 & Nasab, Ali Rezaei, Maedeh Dashti, Mojtaba Shahin, Mansooreh Zahedi, Hourieh Khalajzadeh, Chetan Arora, and Peng Liang. ``A Study of Fairness Concerns in AI-based Mobile App Reviews." arXiv preprint arXiv:2401.08097 (2024).\\
    S37 & Obie, Humphrey O., Hung Du, Kashumi Madampe, Mojtaba Shahin, Idowu Ilekura, John Grundy, Li Li, Jon Whittle, Burak Turhan, and Hourieh Khalajzadeh. ``Automated detection, categorisation and developers’ experience with the violations of honesty in mobile apps." Empirical Software Engineering 28, no. 6 (2023): 134.\\
    S38 & Obie, Humphrey O., Idowu Ilekura, Hung Du, Mojtaba Shahin, John Grundy, Li Li, Jon Whittle, and Burak Turhan. ``On the violation of honesty in mobile apps: Automated detection and categories." In Proceedings of the 19th International Conference on Mining Software Repositories, pp. 321-332. 2022.\\
    S39 & Obie, Humphrey O., Waqar Hussain, Xin Xia, John Grundy, Li Li, Burak Turhan, Jon Whittle, and Mojtaba Shahin. ``A first look at human values-violation in app reviews." In 2021 IEEE/ACM 43rd International Conference on Software Engineering: Software Engineering in Society (ICSE-SEIS), pp. 29-38. IEEE, 2021.\\
    S40 & Mukherjee, Debjyoti, Alireza Ahmadi, Maryam Vahdat Pour, and Joel Reardon. ``An empirical study on user reviews targeting mobile apps’ security \& privacy." arXiv preprint arXiv:2010.06371 (2020).\\
    S41 & Shahin, Mojtaba, Mansooreh Zahedi, Hourieh Khalajzadeh, and Ali Rezaei Nasab. ``A study of gender discussions in mobile apps." In 2023 IEEE/ACM 20th International Conference on Mining Software Repositories (MSR), pp. 598-610. IEEE, 2023.\\
    S42 & Khalajzadeh, Hourieh, Mojtaba Shahin, Humphrey O. Obie, Pragya Agrawal, and John Grundy. ``Supporting developers in addressing human-centric issues in mobile apps." IEEE Transactions on Software Engineering 49, no. 4 (2022): 2149-2168.\\
    S43 & Zhang, Jianzhang, Jinping Hua, Nan Niu, Sisi Chen, Juha Savolainen, and Chuang Liu. ``Exploring privacy requirements gap between developers and end users." Information and Software Technology 154 (2023): 107090.\\
    S44 & Dos Santos, Paulo Sérgio Henrique, Alberto Dumont Alves Oliveira, Thais Bonjorni Nobre De Jesus, Wajdi Aljedaani, and Marcelo Medeiros Eler. ``Evolution may come with a price: analyzing user reviews to understand the impact of updates on mobile apps accessibility." In Proceedings of the XXII Brazilian Symposium on Human Factors in Computing Systems, pp. 1-11. 2023.\\
    S45 & Besmer, Andrew R., Jason Watson, and M. Shane Banks. ``Investigating user perceptions of mobile app privacy: An analysis of user-submitted app reviews." International Journal of Information Security and Privacy (IJISP) 14, no. 4 (2020): 74-91.\\
    S46 & Reyes Arias, Jose E., Kale Kurtzhall, Di Pham, Mohamed Wiem Mkaouer, and Yasmine N. Elglaly. ``Accessibility feedback in mobile application reviews: A dataset of reviews and accessibility guidelines." In CHI Conference on Human Factors in Computing Systems Extended Abstracts, pp. 1-7. 2022.\\
    S47 & Harkous, Hamza, Sai Teja Peddinti, Rishabh Khandelwal, Animesh Srivastava, and Nina Taft. ``Hark: A deep learning system for navigating privacy feedback at scale." In 2022 IEEE Symposium on Security and Privacy (SP), pp. 2469-2486. IEEE, 2022.\\
    \hline
\end{longtblr}

\newpage

\section{Data extracted from primary studies} \label{appendixB}

\begin{longtblr}[
    caption={List of the primary studies and data extracted from them},
    label={tab:data}
]{
  colspec = {lp{7cm}p{4.5cm}p{2.5cm}},
  rowhead = 1,
}
    \hline
    \textbf{Study} & \textbf{Data source} & \textbf{Software category} & \textbf{User reviews dataset} \\
    \hline
    \\
    S1 & Google Play (40 apps) \& Apple App (16 apps) & Depression & 2217 reviews \\
    S2 & Meta Quest (1 app - Rec Room) & Social VR & 1000 reviews \\
    S3 & Google Play (12783 apps) & General & 13129783 reviews \\
    S4 & Google Play \& Apple App stores (18 apps) & Investing, Food delivery, Mental health & 940630 reviews \\
    S5 & Reddit & General & 4488467 reviews \\
    S6 & Google Play (117 apps) \& Apple App (76 apps) & Mental health & 4923 reviews \\
    S7 & Google Play (50 apps) & Mobile payment & 1886352 reviews \\
    S8 & Google Play (8999 apps) & Gaming & 56439878 reviews \\
    S9 & Twitter & Sharing economy & 667806 reviews \\
    S10 & Google Play (41 apps) & Mental health & 41000 reviews \\
    S11 & Google Play (3 apps) & Metaverse & 120000 reviews \\
    S12 & Google Play (8 apps) \& Apple App (8 apps) & Teletherapy & 3268 reviews \\
    S13 & Google App \& Apple App (278 apps) & Fitness \& Health & 5870851 reviews \\
    S14 & Google Play (2583 apps) & General & 4547493 reviews \\
    S15 & Google Play \& Steam (50 apps) & Gaming & 833 reviews \\
    S16 & Apple App (1 app) & Lodging sharing economy & 500 reviews \\
    S17 & Google Play (2M apps) & General & 580M reviews \\
    S18 & Google Play (340 apps) & General & 179519598 reviews \\
    S19 & Google Play (701 apps) & General & 5326 reviews \\
    S20 & Google Play (6938 apps) & General & 5108538 reviews \\
    S21 & Google Play (200 apps) & General & 812899 reviews \\
    S22 & Google Play (50 apps) & Mobile payment & 1886352 reviews \\
    S23 & Chrome Web Store (8 extensions) & Education & 613 reviews \\
    S24 & Google Play (17 apps) & General & 64783 reviews \\
    S25 & Google Play (701 apps) & General & 214053 reviews \\
    S26 & Google Play (809 apps) & General & 495051 reviews \\
    S27 & Google Play (15124 apps) & General & 18326624 reviews \\
    S28 & Google Play \& Apple App (30 apps) & COVID-19 & 225708 reviews \\
    S29 & Reddit (12 apps) & Social network & 459523 reviews \\
    S30 & Google Play (183 apps) \& Apple App (254 apps) & Mental health & 101715 reviews \\
    S31 & Google Play (10 apps) & Fitness \& Health & 44643 reviews \\
    S32 & Google Play (10 apps) & General & 3101 reviews \\
    S33 & Google Play, Twitter \& Reddit (50 apps) & Business, Entertainment, Finance, E-Commerce \& Social Network & 10.22M reviews \\
    S34 & Reddit \& DPReview (online forum) & Graphic design & 2666 reviews \\
    S35 & Apple App (10 apps) & Social network & 1886838 reviews \\
    S36 & Google Play (108 apps) & General & 17968298 reviews \\
    S37 & Google Play (713 apps) & General & 236660 reviews \\
    S38 & Google Play (713 apps) & General & 236660 reviews \\
    S39 & Google Play (12 apps) & General & 22607 reviews \\
    S40 & Google Play (539 apps) & General & 2186093 reviews \\
    S41 & Google Play (70 apps) & General & 7M reviews \\
    S42 & Google Play \& GitHub (12 apps) & General & 2400 reviews \\
    S43 & Apple App (1 app) & Social network & 112826 reviews \\
    S44 & Google Play (340 apps) & General & 4999 reviews \\
    S45 & Amazon Appstore (384365 apps) & General & 5033376 reviews \\
    S46 & Google Play (701 apps) & General & 2663 reviews \\
    S47 & Google Play (1.3M apps) & General & 626M reviews \\
    \hline
\end{longtblr}

\begin{longtblr}[
    caption={List of techniques employed by primary studies for identifying and analyzing user reviews.},
    label={tab:methods}
]{
  colspec = {lp{5.5cm}p{3.5cm}p{4.2cm}},
  rowhead = 1,
}
    \hline
    \textbf{Study} & \textbf{Identification of relevant reviews} & \textbf{Automated classification of reviews} & \textbf{Analysis of reviews} \\
    \hline
    \hline
    S1 & Most relevant and most critical filter of app stores & & Thematic analysis \\
    S2 & Manual analysis & & Thematic analysis \\
    S3 & Keyword search and feature augmentation with an initial set (security, privacy) & Sparse SVM &  \\
    S4 & Keyword search and manual analysis & & Hybrid TF-IDF with GloVe \\
    S5 & Keyword search and manual analysis & USE & K-means with Distil-BERT and UMAP \\
    S6 S10 S12 & Most recent filter of app stores and manual analysis & & Thematic analysis \\
    S7 & Keyword search and sentiment analysis & & Thematic analysis, MobSF, AndroBugs, RiskDroid, Microsoft's Accessibility Insights Tool \\
    S8 & Manual analysis & CNN & \\
    S9 & Keyword search and manual analysis & & Conceptual crowd-driver model \\
    S11 & Star rating three or less & & LDA \\
    S13 & BoW and manual analysis & Keyword classification & \\
    S14 & Keyword search & SVM & \\
    S15 & Star rating one or negative filter & & Quantitative content analysis \\
    S16 & Star rating one and two & & LDA \\
    S17 & Regex of n-gram keywords & USE, BERT, BERT-SST, Ensemble model & K-means \\
    S18 & Keyword search and manual analysis & & BERTopic \\
    S19 & Data from S25 & LR, DF, BDTs, NN, SVM, AP, BPM, DJ, LDSVM & \\
    S20 & Keyword search and manual analysis & ANN & \\
    S21 & Keyword search and cosine similarity & SVM, LR, DT, RF, KNN & \\
    S22 & Keyword search and sentiment analysis & & Thematic analysis \\
    S23 & Manual analysis & & Manual analysis \\
    S24 & Keyword search and SDG & & LR \\
    S25 & Keyword search and manual analysis & & Manual analysis \\
    S26 & Keyword search and manual analysis & Semantic rule-based classification & BTM \\
    S27 & Keyword search and manual analysis & SVM, NB, CART, ME & \\
    S28 & Keyword search and manual analysis & & Manual analysis \\
    S29 & Keyword search and manual analysis & LR, SVM, LinearSVC, NuSVC, GNB & \\
    S30 & Manual analysis & & Thematic analysis \\
    S31 & LR from S21 & & Manual analysis \\
    S32 & Manual analysis & LR, MNB, RF, MLP, SVM, BERT & \\
    S33 & Random sampling & GPT-2, BERT, RoBERTa, XLM-RoBERTa, BART & \\
    S34 & BERTopic and manual analysis & & Thematic analysis \\
    S35 & Cosine similarity with all-mpnet, all-MiniLM, multi-qa-mpnet, multi-qa-MiniLM & LR, LinearSVM, RF, GBT, fastText, BERT & BERTopic with K-means \\
    S36 & Keyword search with KeyBERT and manual analysis & LR, SVM, XGB, BERT, RoBERTa, GPT-3 & K-means \\
    S37 & Keyword search and manual analysis & LR, SVM, DT, GBT, NN, DNN, GAN & \\
    S38 & Keyword search and manual analysis & LR, SVM, DT, GBT, NN & \\
    S39 & Keyword search and sentiment analysis & & Manual analysis \\
    S40 & Keyword search and manual analysis & NB, KNN, SLP, SVM & Manual analysis \\
    S41 & Keyword search with KeyBERT and manual analysis & LR, SVM-LR, RF, XGB, AdaBoost, DT, CNN, LSTM, RoBERTa & Manual analysis \\
    S42 & Manual analysis & LR, SVM, RF, XGBoost, BERT, RoBERTa, DistilBERT & \\
    S43 & Cosine similarity with SentenceTransformers & & LDA \\
    S44 & Keyword search and manual analysis & & ChatGPT-4 \\
    S45 & Random sampling, keyword sampling, corrective sampling and manual analysis & LR & Manual analysis \\
    S46 & Keyword search & & Deductive coding \\
    S47 & NLI and manual analysis & T5-11B & T5-11B with manual analysis \\
    \hline
\end{longtblr}

\begin{longtblr}[
    caption={Authors Affiliation},
    label={tab:aff}
]{
  colspec = {lp{11cm}},
  rowhead = 1,
}
        \\
        \hline
        \textbf{Authors Affiliation} & \textbf{Studies} \\
        \hline
        \hline
        Collaboration & S1 S3 S37 S38 S39 S43 \\
        Academic & S2 S4 S5 S6 S7 S8 S9 S10 S11 S12 S13 S14 S15 S16 S18 S19 S20 S21 S22 S23 S24 S25 S26 S27 S28 S29 S30 S31 S32 S33 S34 S35 S36 S40 S41 S42 S44 S45 S46 \\
        Industrial & S17 S47 \\
        \hline
\end{longtblr}

\newpage
\section{Publication Trends} \label{appendixC}

\begin{longtblr}[
    caption={List of primary studies over publication venues and venue types},
    label={tab:venue}
]{
  colspec = {|p{8.5cm}|p{2cm}|p{2cm}|p{1cm}|},
  rowhead = 1,
}
    \hline
    \textbf{Publication venue} & \textbf{Venue type} & \textbf{Studies} & \textbf{No.} \\
    \hline
    CHI Conference on Human Factors in Computing Systems & Conference &	S1, S15, S18, S19, S46 & 5 \\
    \hline
    International Requirements Engineering Conference &	Conference & S5, S9 & 2 \\
    \hline
    IEEE Symposium on Security and Privacy & Conference & S14, S47 & 2 \\
    \hline
    International Conference on Software Engineering & Conference &	S17, S26 & 2 \\
    \hline
    Brazilian Symposium on Human Factors in Computing Systems &	Conference & S25, S44 &	2 \\
    \hline
    International Conference on Mining Software Repositories & Conference &	S38, S41 & 2 \\
    \hline
    International Conference on Mobile Software Engineering and Systems & Conference & S27, S28 &	2 \\
    \hline
    International Conference on Software Engineering: Software Engineering in Society (ICSE-SEIS) &	Conference & S29, S39 &	2 \\
    \hline
    International Conference on Mobile and Ubiquitous Multimedia & Conference &	S2 & 1 \\
    \hline
    ACM SIGSAC Conference on Computer and Communications Security &	Conference & S3 &	1 \\
    \hline
    International Conference on Automated Software Engineering & Conference & S4 & 1 \\
    \hline
    European Symposium on Usable Security &	Conference & S7 &	1 \\
    \hline
    International Joint Conference on Pervasive and Ubiquitous Computing and International Symposium on Wearable Computers & Conference & S8 & 1 \\
    \hline
    International Conference on Data Intelligence and Security & Conference & S16 & 1 \\
    \hline
    International Carnahan Conference on Security Technology & Conference & S20 &	1 \\
    \hline
    Symposium on Usable Privacy and Security (SOUPS) & Conference & S23 & 1 \\
    \hline
    Annual Privacy Forum & Conference & S31 & 1 \\
    \hline
    ACM on Human-Computer Interaction & Journal & S6, S10 & 2 \\
    \hline
    Empirical Software Engineering & Journal & S13, S37 & 2 \\
    \hline
    Information and Software Technology & Journal & S24, S43 & 2 \\
    \hline
    Entertainment Computing & Journal & S11 & 1 \\
    \hline
    JMIR Mental Health & Journal & S12 & 1 \\
    \hline
    Computers \& Security & Journal & S21 & 1 \\
    \hline
    IEEE Access & Journal & S30 & 1 \\
    \hline
    IEEE Transactions on Software Engineering & Journal & S42 & 1 \\
    \hline
    International Journal of Information Security and Privacy (IJISP) & Journal & S45 & 1 \\
    \hline
    Workshop on Privacy in the Electronic Society & Workshop & S22 & 1 \\
    \hline
    arXiv preprint & pre-publications & S32, S33, S34, S35, S36, S40 & 6 \\
    \hline
\end{longtblr}

\newpage
\section{Additional Information} \label{appendixD}

\begin{longtblr}[
    caption={Open source data from primary studies},
    label={tab:datalinks}
]{
  colspec = {l|p{12.5cm}},
  rowhead = 1,
}
    \textbf{Studies} & \textbf{Datasource Link} \\
    \hline
    \hline
    S4 & \url{https://seel.cse.lsu.edu/data/ase22.zip} \\
    \hline
    S5 & \url{https://zenodo.org/records/6272629} \\
    \hline
    S8 & \url{https://ocslab.hksecurity.net/Datasets/review-with-security-concern-dataset} \\
    \hline
    S18 & \url{https://wajdialjedaani.com/A11yChi23/} \\
    \hline
    S24 & \url{https://github.com/christinaghj/SRR-Miner} \\
    \hline
    S25 & \url{https://github.com/marceloeler/data-ihc2019} \\
    \hline
    S26 & \url{https://github.com/UBCFinder/UBCFinder} \\
    \hline
    S27 & \url{http://cs.gssi.it/MobileSoft2018ReplicationPackage} \\
    \hline
    S29 & \url{https://zenodo.org/records/7194259} \\
    \hline
    S32 & \url{https://zenodo.org/records/7755689} \\
    \hline
    S33 & \url{https://zenodo.org/records/10050673} \\
    \hline
    S35 & \url{https://github.com/zhangjianzhang/PCTD} \\
    \hline
    S36 & \url{https://zenodo.org/records/12180182} \\
    \hline
    \SetCell[r=2]{} S37 S38 & \url{https://anonymous.4open.science/r/ml_app_reviews-3ED6/README.md}\\ & \url{https://github.com/kashumi-m/ReplicationPackageMobileAppsHonestyViolations}\\
    \hline
    S41 & \url{https://zenodo.org/records/8238376} \\
    \hline
    S42 & \url{https://zenodo.org/records/6982529} \\
    \hline
    S43 & \url{https://github.com/zhangjianzhang/privacy_re_gap} \\
    \hline
    S45 & \url{https://github.com/WinthropUniversity/MobileAppReviewPrivacy} \\
    \hline
    S46 & \url{https://github.com/thekindlab/Dataset_Apps_Reviews_by_Accessibility_Guidelines} \\
    \hline
    S47 & \url{https://github.com/google/hark}
\end{longtblr}

\begin{longtblr}[
    caption={Example of user reviews from primary studies for each ethical concern},
    label={tab:reviews}
]{
  colspec = {lp{12.5cm}},
  rowhead = 1,
}
    \textbf{Concern} & \textbf{User Reviews} \\
    \hline
    \SetCell[r=1]{} Privacy & \textbf{User Review 1}: Invasive always listening, rarely does what I want, often has inappropriate responses to TV dialogue. Stay away you're wasting your money. You're also giving up a lot of your privacy. \\ & \textbf{User Review 2}: Accesses a bunch of personal information on your phone for no good reason. Pay for the service twice. Once with your privacy and then with your money. \\ & \textbf{User Review 3}: App monitors your texts and calls. Uses app to exploit your personal information. Download at own risk. \\ & \textbf{User Review 4}: It is pathetic to elucidate that [the app] is fast becoming an app where sharing your file is unsafe. The pdf files I shared while holding a meeting got completely lost, nowhere to be found in my tablet again. Something drastic must be done to savage this situation else... \\
    \hline
    \SetCell[r=1]{} Security & \textbf{User Review 1}: I think it should be more secure because my [the app] account has been hacked and [the app] didn't help to stop hacking my account \\ & \textbf{User Review 2}: [the app] was hacked and user's identities have been compromised. I suggest not using this service if you value your privacy. \\ & \textbf{User Review 3}: Govt. of India declare is unsafe also found out that hackers to sell [the app] video app data for approx 23 Lakh that around 29,987USD this app is not safe at all, I am terminating my account right away hope they didn't sell my data already \#boycott[the app] \\ & \textbf{User Review 4}: In this app , the is getting hacked easily and ....anyone unknown can enter into the meeting simultaneously which is very dangerous for our privacy \\
    \hline
    \SetCell[r=1]{} Accessibility & \textbf{User Review 1}: I have a disability and [the app] helps me get to doctor visits. I was disappointed when my service dog was denied a ride with me. I tried filing a complaint, but got no resolution. In fact I was told that drivers, here in Albuquerque, can refuse to give people-even if they have a service dog-a ride? \\ & \textbf{User Review 2}: Could be excellent but no automatic caption facility for the free version of [the app]. Where is the accessibility for Deaf users?? \\ & \textbf{User Review 3}: please give dark mode for this app it’s so eye straining to watch in meet. \\
    \hline
    \SetCell[r=1]{} Accountability & \textbf{User Review 1}: No insurance for back up recordings!!! I am on a free plan given I don't need more than one person per meeting. My computer broke had to replace it hoping I didn't loose my recordings... well [the app] has them, but I can't access them from a different computer. When I tried contacting support you are given a Bot that isn't helpful at all for such a common issue of computer replacements and when I called the number they said I'm not eligible for person to person support... so thank you [the app]. \\ & \textbf{User Review 2}: I just want to complain about something. So I have decided to change my name into my legit name (school purposes) and until now, I still can't access my account. I've already sent a photo of my ID (2), and this has been happening for more than a week now. I hope you'll take actions towards this problem. As soon as possible please. I need my account for school, thank you. \\ & \textbf{User Review 3}: Awful customer service. If you ever have a problem with you item just know that you are just wasting your time contacting customer support. Great concept but awful staff. \\
    \hline
    \SetCell[r=1]{} Fairness & \textbf{User Review 1}: If you're a black creator this app discriminates against you. They allow white supremacists to post hate content but block black creators and their content. If this has happened to you file a lawsuit for discrimination. \\ & \textbf{User Review 2}: Racist against Black and Brown Conservatives and Independents. The left are allowed to threaten and harass POC as well as whites who have different opinions. Posts and comments censored over non violations yet animal abuse and child porn are allowed. Tells you exactly what the left is about. \\ & \textbf{User Review 3}: In India very limited features are offered. It is pure example of discrimination. In US there are so many more features like energy dashboard, hunch, etc... We in India pay the same amount of money for your echo devices and thus deserve all the features... \\
    \hline
    \SetCell[r=1]{} Transparency & \textbf{User Review 1}: App is not very user friendly, signs you up for things that autobill and the cancellation pages are hidden. Also their help menus tell you steps to take that simply are not shown their. They need to actually beta test this thing more. and there is no phone number for customer service, which might be lean for operating costs but its not very professional. Its not at all as self expanatory as their fake news marketing would have you believe. They need to get with the program and improve their app. \\ & \textbf{User Review 2}: You can sell clothes without a fee, which is nice. However, if you use a picture from the internet for example it will delete your post. There are 3 different shipping options based on how the sellers list it. Buyer pays a fee which IS NOT noted anywhere. It is a small fee. The set up is terrible but doable. \\ & \textbf{User Review 3}: Why so many permissions!!! And not a convincing explanation. \\ & \textbf{User Review 4}: Suddenly [the app] stopped views on my videos. In spite of having 38k followers I’m not getting 200 views ... Best videos get no views while 3rd class videos have views in millions. \\
    \hline
    \SetCell[r=1]{} Safety & \textbf{User Review 1}: Poor algorithms that block legitimate content but leave inappropriate and extremist political content up. Its unsafe for the under 18yrs old user. \\ & \textbf{User Review 2}: I love it but need to protect the American people who used this social media to reach out to family friends work and ect so it a lot of fake jobs advertisement on [the app] and using [the app] to steal valuable information about the American people who using [the app] stop the con artist \\ & \textbf{User Review 3}: My [the app] driver saw me waiting, sped up, and hit me putting me in the hospital. To top it off he charged me a cancellation fee. Never again!!! \\ & \textbf{User Review 4}: It's fine, I guess, but their reporting needs work. Like they declined to remove a video of someone giving an instructional on how to make homemade infant formula which is not at all safe. The misinformation is really dangerous sometimes and [the app] refuses to act. \\
    \hline
    \SetCell[r=1]{} Social Solidarity & \textbf{User Review 1}: It is a bad and horrible app that causes sleep ammonia and many more, the app causes so many bad addictions. \\ & \textbf{User Review 2}: This app is not useful and people are wasting there time. It is unethical that [the app] is deleting valid negative reviews. \\ & \textbf{User Review 3}: [the app] is no longer a social platform of free speech and free thinking. [the app] is now just another corrupt arm of the socialist left. Dropping [the app] and heading over to MeWe to enjoy uncensored free speech ... \\ & \textbf{User Review 4}: Extremely Low Quality Ads. Promote prostitution, gambling, illegal drugs. Useless target group.. any person between 20+ to 60 will receive above categories. [the app] ads only can income don't xare what are the contents of ads. \\ 
    \hline
\end{longtblr}

\end{appendices}

\EOD

\end{document}